\documentclass[bibyear]{aa}  
\usepackage{txfonts}
\usepackage{dsfont}
\usepackage{mathtools}
\usepackage{graphicx}
\usepackage{multirow}
\usepackage[authoryear]{natbib}
\usepackage{tabu}

\usepackage[breaklinks=true]{hyperref}

\makeatletter
\providecommand*{\diff}%
{\@ifnextchar^{\DIfF}{\DIfF^{}}}
\def\DIfF^#1{%
\mathop{\mathrm{\mathstrut d}}%
\nolimits^{#1}\gobblespace
}
\def\gobblespace{%
\futurelet\diffarg\opspace}
\def\opspace{%
\let\DiffSpace\!%
\ifx\diffarg(%
\let\DiffSpace\relax
\else
\ifx\diffarg\[%
\let\DiffSpace\relax
\else
\ifx\diffarg\{%
\let\DiffSpace\relax
\fi\fi\fi\DiffSpace}
\newcommand{\subrm}[2]{#1_{\rm #2}}
\newcommand{\difp}[2]{\frac{\partial #1}{\partial#2}}
\newcommand{\vect}[1]{\mathbf{#1}}

\begin{document}

   \title{Evolution of angular-momentum-losing exoplanetary systems}

   \subtitle{Revisiting Darwin stability}

   \author{   C.~Damiani\inst{1}
                   \and
                   A.~F.~Lanza\inst{2}
              }

   \institute{   Aix-Marseille Universit{\'e}, CNRS, Laboratoire d\rq{}Astrophysique de Marseille, UMR 7326, 13388, Marseille, France\\
                     \email{cilia.damiani@lam.fr}
                     \and
                      INAF - Osservatorio Astrofisico di Catania, Via Santa Sofia 78, 95123, Catania, Italy
                 }

   \date{Received xxx; accepted xxx}
 
  \abstract
  {}
  {We assess the importance of tidal evolution and its interplay with magnetic braking in the population of hot-Jupiter planetary systems.}
  {By minimizing the total mechanical energy of a given system under the constraint of stellar angular momentum loss, we rigorously find the conditions for the existence of dynamical equilibrium states. We estimate their duration, in particular when the wind torque spinning down the star is almost compensated for by the tidal torque spinning it up. We introduce dimensionless variables to characterize the tidal evolution of observed hot Jupiter systems and  discuss their  spin and orbital states using generalized Darwin diagrams based on our new approach.}
 {We show that their orbital properties are  related to the effective temperature of their host stars. The long-term evolution of planets orbiting F- and G-type stars is significantly different owing to the combined effect of magnetic braking and tidal dissipation. The existence of a quasi-stationary state, in the case of short-period planets, can significantly delay their tidal evolution that would otherwise bring the planet to fall into its host star. Most of the planets known to orbit F-type stars are presently found to be near this stationary state,  probably in a configuration not too far from what they had when their host star settled on the zero-age main sequence.  Considering the importance of angular momentum loss in the early stages of stellar evolution, our results indicate that it has to be considered to properly test the migration scenarios of planetary system formation.}
   {}


   \maketitle
\section{Introduction}
\label{sec:intro}
Over the past two decades, detection and characterization of hundreds of exoplanets has revealed an unexpectedly broad diversity of planets and orbital configurations. Among the various detection methods, the radial velocity and transit techniques, two indirect methods, have so far been the most successful.  When they are combined, they allow one to infer the masses and radii of the star and planet up to a one-parameter degeneracy \citep[see e.g.][]{Wright2013}. The easiest planets to detect and fully characterize  are those with a mass comparable to that of Jupiter in close-in orbits ($\leq 0.1$~AU) around  main-sequence stars. They form a significant proportion of the known exoplanets, usually called the \lq\lq{}hot Jupiters\rq\rq{}. For these planets, the orbital, planetary, and stellar host main parameters can be precisely and accurately determined. Many of their observed properties, however, still have to be understood, such as the \lq\lq{}radius anomaly\rq\rq{} \citep{Guillot2006,Laughlin2011}, the origin of the observed eccentricities \citep{Ford2008}, obliquities \citep{Triaud2011}, or the very fact that they are orbiting so close to their host stars. Those properties can result directly from the formation processes or could have been acquired during the evolutionary lifetime of the system.

According to the prevailing theory \citep{Pollack1996, Mordasini2008}, giant planets are formed within a protoplanetary disk, and require a solid core to first be assembled to allow efficient subsequent capture and growth within the relatively short disk lifetime \citep[$\lesssim 5$~Myr for $\sim 50$~percent of the protostars, ][]{Mamajek2009}. This implies that giant planets must form beyond the snow line located typically at a few astronomical units from the star. Hot Jupiters have a semi-major axis $a \lesssim 0.1$ AU, so they must have undergone some kind of migration. Two main mechanisms have been proposed: either migration occurs within the protoplanetary disk and involves torques between the planet and the surrounding gas \citep{Lin1996}, or alternatively, it can be the result of dynamical instabilities associated with the gravitational interactions among two or more bodies orbiting the star  after the evaporation of the disk \citep{Rasio1996}. Those migration theories involve different halting mechanisms that can be tested by comparing their  predictions with the observed orbital properties of exoplanets \citep{Plavchan2013}.

Further secular changes in the orbits of exoplanets can still be induced by tidal interaction between the planet and the star, even when the primordial migration mechanism is no longer effective. The tidal torque scales as the inverse of the sixth power of the semi-major axis $a^{-6}$, consequently it is especially important in the case of hot Jupiters. To test the migration scenarios, it is thus crucial to estimate the efficiency of tidal dissipation and its effects over the evolutionary lifetime of the star. We do not engage in an exhaustive review but rather recall the three major limitations for that kind of study: the knowledge of the actual mechanism responsible for tidal dissipation, its efficiency, and the effects of the loss of angular momentum of the system through the magnetic braking of the host.

In the present work, we first review our current knowledge of the processes ruling the evolution of the angular momentum in a planetary system, considering both tides and stellar magnetic braking (Sect.~\ref{theorysect}). A new general discussion of the equilibrium configurations that a system can attain during its tidal evolution including stellar magnetic braking is introduced in Sect.~\ref{stabGen}. Then we apply our theory to a sample of planetary systems  and discuss their evolution using a particularly simple graphic approach that generalizes the classic Darwin tidal diagrams (Sects.~\ref{stabHJ} and ~\ref{SecEvol}). Finally, we discuss the implications of our results for tidal dissipation efficiency in late-type stars and for the mechanisms of formation and evolution of planetary systems (Sect.~\ref{discussion}). 

\section{Tides and angular momentum evolution in late-type stars}
\label{theorysect}
\subsection{Tidal dissipation theories}\label{tidtheo}
The response of a fluid body to tidal forcing can be separated into two components: the equilibrium tide that represents a large-scale, quasi-hydrostatic distortion of the body, and the dynamical tide, which corresponds to the response of the oscillation modes that are excited by the time-dependent tidal potential. Since the first attempts to derive a theory of tides in a fluid body \citep{Zahn1966a,Zahn1966b,Zahn1966c}, the main difficulty has been to identify the physical processes that are actually responsible for the conversion of the tidal torque mechanical energy into heat. While turbulent viscosity acting on the equilibrium tide can successfully reproduce the circularization of stars possessing a large convective envelope \citep{Verbunt1995, Zahn1989}, it fails to provide sufficient dissipation when the convective turnover timescale is much longer than the tidal period, which is usually the case for gaseous planets or low-mass main-sequence stars  \citep{Goodman1997}. For short-period planets, this reduced efficiency of the turbulent viscosity would imply circularization times that are considerably longer than the ages of their host stars \citep{Ogilvie2004}. On the other hand, the development of the dynamical tide theory, including the effects of the Coriolis force, stellar evolution, magnetic braking, and resonance locking \citep{Savonije2002, Witte2002, Ogilvie2004}, has improved the estimation of the efficiency of dissipation in the case of solar-type stars, but the details of wave excitation and damping have not yet been fully understood \citep{GoodmanLackner09}. 

The efficiency of tidal dissipation is usually parametrized by the dimensionless quality factor $Q$ proportional to the ratio of the total kinetic energy of the tidal distortion to the energy dissipated in one tidal period \citep[e.g. ][]{Zahn2008}. It is convenient to introduce the reduced quality factor $Q^{\prime} \equiv (3/2) (Q/k_{2})$,  where $k_{2}$ is the Love number of the body and it measures its density stratification, so that $Q^{\prime} = Q$ for a homogeneous body without rigidity (for which $k_2=3/2$)\footnote{Note that $k_{2}$ is twice the apsidal motion constant of the star, often indicated with the same symbol as in e.g. \citet{Claret1995}.}. A lower value of $Q\rq{}$ implies a stronger tidal dissipation. While some authors have treated $Q\rq{}$ as a constant \citep{Goldreich1966, Ferraz2008, Jackson2008, Jackson2009}, many studies have stressed the importance of including the  dependence of $Q\rq{}$ on the tidal frequency, both for the equilibrium tide \citep{Goldreich1977, Goodman1997,  Leconte2010, Penev2011, Remus2012} and  the dynamical tide \citep{Ogilvie2004, Ogilvie2007}.  Specifically, \citet{Ogilvie2007} have shown in the framework of the dynamical tide that for solar-type stars,  the value of $Q^{\prime}$ decreases by two to four orders of magnitude when the orbital frequency becomes  less than twice the stellar rotational frequency, because tidal dissipation in the convective zone is substantially enhanced by the excitation of inertial waves. More massive stars do not experience as much frequency dependence because the Coriolis force has little net effect due to their thin convective envelope. 

Another consequence of the strong effect of the dissipation of inertial waves is that the average value of $Q\rq{}$ is expected to be  greater for F-type stars than for G-type stars for a given rotation rate because the former have a shallower convection zone than the latter. Moreover, since the mass of the outer convection zone decreases rapidly with increasing stellar mass among F-type stars, the average value of $Q^{\prime}$ is expected to increase by three to four orders of magnitude when the mass ranges between $1.2$ and $1.5$~M$_{\odot}$ \citep{Barker2009}. Tidal dissipation efficiency thus strongly depends on the extension of the outer convective zone, but also on the rotational evolution of the star for which a quantitative global theory is still needed, as discussed in the next section.

\subsection{Evolution of the rotation of late-type stars}\label{magbrak}
The observed rotational period of stars show a clear, although not simple dependence with stellar mass and age \citep[see e.g.][]{Kraft1967, Irwin2009}. While early-type stars remain fast rotators until the end of the main-sequence, F-, G-, and K-type stars have a mean rotation velocity that decreases in time. It is now generally admitted that the convective zone of late-type stars host a hydromagnetic dynamo at the origin of their magnetic activity, which is in turn responsible for the angular momentum loss (AML). This is generally explained by magnetic braking, where a magnetized wind can efficiently extract angular momentum from the star with a very low mass loss rate. In a simplified formulation \citep[see e.g. ][]{Kawaler1988}, the wind torque can be computed considering that the charged particles of the wind follow the field lines of the corona that are frozen in the plasma and rotate with the star as if it were a solid body. The angular momentum is then extracted from the system at a radial distance $r_{\rm A}$ where the wind velocity equals the Alfv\'{e}n velocity. 

How the Alfv\'{e}n radius $r_{\rm A}$ depends on the mass, radius, magnetic field strength and rotation speed is currently not very well known. Reliable computations require knowledge of the wind acceleration profile and the magnetic field geometry above the surface of the star, which remains a challenge. Using a dipole field geometry in MHD simulations, \citet{Matt2012} computed the mass loss rate expected for different values of rotational speed and magnetic field strength, and found significant differences with the usual analytic  prescriptions, such as those  by \citet{Kawaler1988}, especially concerning the dependence of $r_{\rm A}$ on the stellar parameters. Moreover, it may depend on  the geometry of the magnetic field, as shown in \citet{Matt2008} in the case of a pure quadrupolar field. It is, however, not yet clear how more complex magnetic configurations will change the scaling \citep{Pinto2011}. Recent observations indeed show a wide variety in the basic properties of stellar magnetic fields \citep{Donati2009}, with very different field strengths, configurations and degree of axi-symmetry, implying possibly different braking laws with complex dependences on the stellar parameters and rotation rate. 

Another characteristic of the rotation of low-mass stars is that there is a wide spread in rotation periods at different ages \citep{Gallet2013}. Interactions with the protoplanetary disk must play a role during the first 5 Myr or so, but the persistence of fast rotators after a few hundred Myr, and the final homogeneous rotation rate at the age of the Sun can only be produced by different braking laws \citep{Irwin2009} between fast and slow rotators. This trend is also observed in short-period late-type binaries \citep{vantVeer1988, vantVeer1989, Maceroni1991}. Initially fast rotators can retain a fast spin on the main sequence if some mechanism induces a saturation of the AML rate beyond some threshold  angular velocity \citep{Barnes1996}. The actual mechanism responsible for saturation has not yet been clearly identified \citep{Cranmer2011}, and the threshold angular velocity ranges between $3$ and $15 \Omega_\odot$ depending on the braking law considered \citep[cf. Table 4 in][]{Gallet2013}, where $\Omega_\odot = 2.85\times 10^{-6}$~s$^{-1}$ is the present rotation rate of the Sun. Moreover, the surface rotation period is also affected by the internal magnetohydrodynamical transport mechanisms that  redistribute angular momentum inside the stars themselves \citep[see e.g.][]{Charbonnel2013}. 

The detailed evolution of the surface rotation thus depends on physics that has neither been modelled nor observed to date, but there are two observational facts that can be considered robustly established. First, the observed specific angular momentum of stars decreases by one (respectively two) order of magnitude for initially slow (respectively fast) rotating solar-type stars between the disappearance of the disk and the age of the Sun. Second, F-type stars lose angular momentum very slowly during main-sequence evolution, and their characteristic spin down time can be as much as 10 to 100 times longer than that of G-type stars \citep{Wolff1997}.

\subsection{Effects of magnetic braking on tidal evolution on short-period planets}\label{effmag}
Most studies on tidal evolution of close-in planets have considered individual systems, and they generally neglect the effect of stellar magnetic braking \citep[e.g. ][]{Patzold2004, Carone2007, Ferraz2011}. However, the orbital angular momentum of hot Jupiters is of the same order of magnitude of the rotation angular momentum of their host stars, and in some cases, magnetic braking has been shown to be essential to describe the past evolution of orbital elements \citep[cf. ][]{Lanza2011}. In a more general approach, \citet{Dobbs-Dixon2004} have considered how the evolution of the spin of the host star can affect the eccentricity of a planetary orbit, and propose that all main-sequence dwarfs attain a quasi-steady equilibrium state in which the host star\rq{}s AML through the stellar wind is balanced by the tidal transfer of angular momentum from their planets. Due to insufficient data at the time, their theory remained conjectural. 

A formulation of the long-term tidal evolution of close-in planets, including dissipation in both the star and planet and the braking torque, has  been proposed by \citet{Barker2009} or \citet{Bolmont2011}. The former emphasize the importance of the coupled evolution of rotational and orbital elements, for it can result in a much faster evolution than simple timescale estimates predict. The latter show that different stellar spin evolutions have an effect on the orbital evolution mainly for giant planets and that close-in planets orbiting initially slow rotators have a significantly shorter lifetime than those around faster rotators. They conclude, however, that differentiating one spin evolution from another, given the present position of planets, can be very tricky and that better estimates of stellar ages  are needed to constrain tidal-dissipation efficiency. As a consequence, there is to date no general description of the observed orbital properties of hot Jupiters as a result of their tidal evolution under the influence of the magnetic braking of their star. As a matter of fact, the total angular momentum of the star-planet system is not conserved in this case and is decreasing with time. This last point presents a major difficulty when trying to infer the initial properties of the orbits, and especially the question of primordial eccentricity, because the problem is not holomic, and the final state depends on the initial conditions, as well as on the path taken. This makes general conclusions on the global properties of the population of known exoplanets impractical, especially considering that the unknown quantities can vary from one  system to the other.

There is a way to assess the general outcome of tidal evolution even when the details of the dissipation mechanism are not known, using energy considerations alone. Indeed, by examining the extrema of the total energy of a binary system under the constraint of conservation of its total angular momentum, \citet{Darwin1879} illustrated with a graphical method that the outcome of tidal evolution can be twofold: either the two bodies spiral in towards each other until one of them reaches the Roche limit or an equilibrium state is reached asymptotically. It has become customary to call the latter systems \lq\lq{}Darwin stable\rq\rq{}. The existence of such a stable equilibrium  depends on the total angular momentum of the system, while its fate  depends on the relative distribution of the total angular momentum between  stellar spins and  orbital motion. Darwin's approach consisted in plotting some quantity related to the spin angular momentum of the system versus another proportional to the orbital angular momentum. We propose  to adopt the same approach but without imposing the conservation of total angular momentum on the system. This is appropriate in the framework of our consideration of magnetic braking, because angular momentum is extracted from the star by the stellar wind. In other words, given that magnetic braking exerts a torque on the star, the total angular momentum of the star-planet system is not conserved. 
\section{Pseudo-stability of tidal equilibrium} \label{stabGen}

\subsection{Characterization of the equilibrium} 
We consider a system formed by a star and a gravitationally bound companion of masses $\subrm{M}{\star} $ and $\subrm{M}{p}$, respectively, and radii $\subrm{R}{\star}$ and $\subrm{R}{p}$. The periodically varying potential experienced by both objects generates a tidal disturbance in the fluid. Regardless of the mechanism, dissipation of the tides is directly associated with the secular transfer of angular momentum between the spin and the orbit, as well as a loss of energy from the system. \citet{Hut1980} used the method of Lagrange multipliers to rigorously prove that, under the constraint of conservation of total angular momentum, the minimum of energy yields only one possible type of equilibrium that is characterized by co-planarity, circularity, and co-rotation. Including magnetic braking in our problem then means that the minimization of the energy is no longer carried out under the constraint of constant angular momentum, but imposing that it shall be some unknown function. To rigorously estimate the magnetic braking, the dependence on the mass loss rate, strength of the magnetic field, stellar radius, surface gravity, and spin rate must be included \citep{Matt2012}.  In the case of the Sun, the mass loss rate is very low ($\dot{M} \sim 10^{-14} M_\odot\ \rm{yr}^{-1}$), we assume that comparable mass-loss rate can be expected for late-type stars on the main-sequence, and we neglect its effect on the moment of inertia and gravitational forces. For the sake of generality, we consider some simplifications (e.g. dipolar magnetic field and field strength proportional to the rotation rate to some power), and we thus assume that magnetic braking depends only on the mass, radius, and rotation rate of the star \cite[as in][]{Kawaler1988}. Thus, for given star (i.e. for $R_{\star}$ and $M_{\star}$ fixed), our optimization is carried out by assuming that the total angular momentum is a function of the stellar angular velocity alone $\vect{f }(\boldsymbol{\Omega}_\star)$, which has continuous first derivative. 

The dynamical state of a binary system can be specified by 12 parameters: here we choose to take the six classical orbital elements together with the angular velocity vectors of the two objects. To investigate the exchange and dissipation of energy and angular momentum, only three orbital elements are relevant: the semi-major axis $a$, the eccentricity $e$, and the angle  $i$ between the orbital angular momentum $\vect{h}$ and the total angular momentum of the binary $\vect{L}$. It is convenient to choose the $z$-axis along $\vect{L}$ and the $x$-axis such that
\begin{align}
\vect{L}= \begin{bmatrix}
       		0\\
       		0\\
      		L
     		\end{bmatrix}    ,		
     		& \qquad 
\vect{h} = \begin{bmatrix}
       		h \sin i\\
       		0\\
      		h \cos i
     		\end{bmatrix}    .				
 \end{align}    	
with $i \in [0, \pi/2[$ and $h=\sqrt{G \frac{\subrm{M}{\star}^2\subrm{M}{p}^2}{\subrm{M}{\star}+\subrm{M}{p}} a (1-e^2)}$, $G$ being the gravitation constant. In this frame of reference, the angular velocity of the star and the planet can be defined by their Cartesian components, i.e.  
 \begin{align}
\boldsymbol{\Omega}_\star = \begin{bmatrix}
       					\subrm{\Omega}{x}\\
       					\subrm{\Omega}{y}\\
      					\subrm{\Omega}{z}
     					\end{bmatrix}  ,  		
     					&\text{ and} \qquad 
 \subrm{\boldsymbol{\omega}}{p} = \begin{bmatrix}
       					\subrm{\omega}{x}\\
       					\subrm{\omega}{y}\\
      					\subrm{\omega}{z}
     					\end{bmatrix}				
 \end{align}    	
where $|\boldsymbol{\Omega}_\star|= \Omega$ and $| \subrm{\boldsymbol{\omega}}{p}| = \omega$. The total angular momentum of the binary can be written as
 \begin{equation}\label{ltotgen}
 \vect{L} = \vect{h} + \subrm{C}{\star}\boldsymbol{\Omega}_\star + \subrm{C}{p} \subrm{\boldsymbol{\omega}}{p},
 \end{equation}
 where $\subrm{C}{p}$ and $C_\star$ denote the moment of inertia about the rotation axis of the planet and the star, which are considered as rigid bodies, respectively. Both those moments can be written as $C= M (\subrm{r}{g}R)^2$ where $\subrm{r}{g}$ is the non-dimensional radius of gyration. The total energy of the system is the sum of the mechanical energy of the orbit and the rotational kinetic energy of the star and the planet: 
 \begin{equation}
E = - G \frac{\subrm{M}{\star}\subrm{M}{p}}{2a} + \frac{1}{2}\subrm{C}{\star} |\boldsymbol{\Omega}_\star|^2+ \frac{1}{2}\subrm{C}{p} | \subrm{\boldsymbol{\omega}}{p}|^2. 
 \end{equation}      
     
Let $\mathbf{x}= (a, e, i, \subrm{\Omega}{x},\subrm{\Omega}{y},\subrm{\Omega}{z},\subrm{\omega}{x},\subrm{\omega}{y},\subrm{\omega}{z})$ be the nonuple of our nine parameters. We want to find the stationary points of the total energy $E(\mathbf{x} )$ subject to the set of constraints $\vect{L}(\mathbf{x}) = \vect{f }(\boldsymbol{\Omega}_\star$) or equivalently $\boldsymbol{\Psi}(\mathbf{x} ) = \vect{L}(\mathbf{x}) - \vect{f }(\boldsymbol{\Omega}_\star$) = 0.  With the chosen $z$-axis, we can write $\vect{f}(\boldsymbol{\Omega}_\star)= (0,0,\subrm{f}{z}(\subrm{\Omega}{x},\subrm{\Omega}{y},\subrm{\Omega}{z}) )$. We introduce the Lagrange function $\Lambda$  defined as
\begin{equation}
\Lambda (\mathbf{x}, \lambda_{\rm x},  \lambda_{\rm y}, \lambda_{\rm z}) =  E(\mathbf{x}) + \sum_{j=1}^3 \lambda_j  \Psi_j(\mathbf{x}),
\end{equation}
where $j$ is the subscript for the $(x,y,z)$ Cartesian coordinates. The stationary points of the energy under the given constraint necessarily satisfy the following condition
\begin{equation}\label{lagmeth}
\nabla \Lambda(\mathbf{x}, \lambda_{\rm x},  \lambda_{\rm y}, \lambda_{\rm z}) = 0.
\end{equation}
Considering the simplifying assumptions regarding the magnetic braking introduced above, we assume that $\subrm{f}{z}$ is a function of only the stellar angular velocity $\boldsymbol{\Omega}$, and we can write the partial derivatives of $\subrm{f}{z}$ for $j\in\{x,y,z\}$ as
\begin{equation}\label{changevar}
\difp{\subrm{f}{z}}{\Omega_j}=\frac{\diff \subrm{f}{z}}{\diff \Omega} \difp{\Omega}{\Omega_j}. 
\end{equation}
 Since we have $ \Omega= \sqrt{\subrm{\Omega}{x}^2+\subrm{\Omega}{y}^2+\subrm{\Omega}{z}}^2$, Eq.~\ref{changevar} is simply
\begin{equation}\label{eqsimder}
\difp{\subrm{f}{z}}{\Omega_j}= \frac{\diff \subrm{f}{z}}{\diff \Omega} \frac{\Omega_j}{\Omega}\equiv f^\prime\frac{\Omega_j}{\Omega}.
\end{equation} 
Using Eq.~\ref{eqsimder} in Eq.~\ref{lagmeth} gives the following system of twelve equations:
\begin{align}
&\frac{G{\rm M}_\star \subrm{M}{p}}{a} + h (\lambda_{\rm x} \sin i + \lambda_{\rm z} \cos i ) = 0 \label{s1}\\
&\frac{e}{1-e^2}h (\lambda_{\rm x} \sin i + \lambda_{\rm z} \cos i )   = 0 \label{s2}\\
&\lambda_{\rm x} \cos i - \lambda_{\rm z}   \sin i = 0 \label{s3}\\
&\subrm{\Omega}{x} ( \subrm{C}{\star}  - \lambda_{\rm z}   \frac{f^\prime} {\Omega} ) + \subrm{C}{\star} \lambda_{\rm x}= 0 \label{s4}\\
&\subrm{\Omega}{y}  ( \subrm{C}{\star}  - \lambda_{\rm z}  \frac{f^\prime} {\Omega}) + \subrm{C}{\star}   \lambda_{\rm y} = 0 \label{s5} \\
&\subrm{\Omega}{z}  ( \subrm{C}{\star}  - \lambda_{\rm z}   \frac{f^\prime} {\Omega} )+  \subrm{C}{\star}  \lambda_z = 0 \label{s6}\\
&\subrm{\omega}{x} + \lambda_{\rm x} = 0 \label{s7}\\
&\subrm{\omega}{y} + \lambda_{\rm y}= 0 \label{s8}\\
&\subrm{\omega}{z} + \lambda_{\rm z} = 0 \label{s9}\\
& h \sin i +  \subrm{C}{\star}  \subrm{\Omega}{x} + \subrm{C}{p}  \subrm{\omega}{x}  =0  \label{s10}\\
&\subrm{C}{\star}  \subrm{\Omega}{y} +  \subrm{C}{p} \subrm{\omega}{y}= 0  \label{s11}\\
& h \cos i + \subrm{C}{\star}  \subrm{\Omega}{z} +  \subrm{C}{p} \subrm{\omega}{z} - \subrm{f}{z} = 0 \label{s12}
\end{align} 

The solutions of this system of twelve equations yield $e~=~i~=~0$ and $\subrm{\Omega}{x}=\subrm{\Omega}{y} = \subrm{\omega}{x} = \subrm{\omega}{y} =0$. The stationary points of the energy are thus characterized by circularization and co-planarity as in the case of conserved angular momentum. On the other hand, the synchronization condition now becomes 
\begin{align}
\omega &= n \label{eq:lomsync}\\
\Omega &= n \bigl( 1- \frac{1}{\subrm{C}{\star}} \frac{\diff \subrm{f}{z}}{\diff \Omega} \bigr)\label{eq:omsync}
\end{align}
where $n$ is the mean orbital motion, and provided that $\frac{1}{\subrm{C}{\star}} \frac{\diff \subrm{f}{z}}{\diff \Omega}\neq1$.  We recover a result similar to the case where magnetic braking is neglected, but the equilibrium is now characterized by the quasi-co-rotation of the stellar spin with the orbital mean motion. Let us define 
\begin{equation}\label{defbeta}
\beta \equiv 1-\frac{1}{\subrm{C}{\star}}\frac{\diff \subrm{f}{z}}{\diff\Omega},
\end{equation}
so the equilibrium is characterized by $\Omega=\beta n$. Now let us examine the possible values of $\beta$. The derivative of the function $\subrm{f}{z}$ with respect to $\Omega$ can be computed, noting that
\begin{equation}
\frac{\diff \subrm{f}{z}}{\diff\Omega} = \frac{\diff L}{\diff t} \left(\frac{\diff \Omega}{\diff t} \right)^{-1}.
\end{equation}
Thus Eq.~\ref{defbeta} can be written as
\begin{equation}\label{eqldot}
 \frac{\diff L}{\diff t} = (1-\beta) \subrm{C}{\star}  \frac{\diff \Omega}{\diff t}.
\end{equation}
\begin{table*}
 \caption{Hessian of the energy at an equilibrium point}
 \centering  
 \label{tab:hess} 
\begin{tabular}{c| c c c c c c| }
\multirow{6}{*}{$H=$}	&$\frac{G\subrm{M}{\star} \subrm{M}{p}}{4 a^3} (\frac{\alpha}{C_\star} + \beta -4 )$ &0 &0 &0&0 &$\frac{G\subrm{M}{\star} \subrm{M}{p}}{2a^2n}\frac{\subrm{C}{p}} {\subrm{C}{\star}}$\\
					&0&$\frac{G\subrm{M}{\star} \subrm{M}{p}}{a}\beta $ &0&0&0&0\\
	 				&0&0&$\frac{G\subrm{M}{\star} \subrm{M}{p}}{a}(\frac{\alpha}{\subrm{C}{\star}}+\beta) $&$\frac{\subrm{C}{p}} {\subrm{C}{\star}} \frac{G\subrm{M}{\star} \subrm{M}{p}}{an}$&0&0\\ 
					&0&0&$\frac{\subrm{C}{p}} {\subrm{C}{\star}} \frac{G\subrm{M}{\star} \subrm{M}{p}}{an}$&$\frac{\subrm{C}{p}}{\subrm{C}{\star}}(\subrm{C}{p} +\subrm{C}{\star})$&0&0\\
	 				&0&0&0&0&$\frac{\subrm{C}{p}}{\subrm{C}{\star}}(\subrm{C}{p} +\subrm{C}{\star})$&0\\
	 				&$\frac{G\subrm{M}{\star} \subrm{M}{p}}{2a^2n}\frac{\subrm{C}{p}} {\subrm{C}{\star}}$&0&0&0&0&$\frac{\subrm{C}{p}}{\subrm{C}{\star}}(\subrm{C}{p} +\subrm{C}{\star})$
\end{tabular}
\end{table*}
The rotation rate of the star is controlled by two torques: a) the magnetized wind torque that can only spin down the star; and b) the tidal torque, that can only spin up the star when $\Omega < n$. The parameter $\beta$ can be seen as the ratio of the tidal torque to the total torque acting on the star. A value $\beta \approx1$ corresponds to the case where the total angular momentum of the system is approximately conserved and  is equivalent to the case where magnetic braking is neglected. A value  $\beta \approx0$ corresponds to the case where the total AML of the system is the AML of the star where the tidal torque is negligible. Since the total angular momentum of the system can only decrease so that $\dot{L} \leq 0$ at all times, we immediately see that if $\beta \leq 1$, then necessarily $\dot{\Omega}\leq 0$, and if $\beta \geq 1$ then $\dot{\Omega}\geq 0$. In this way, whenever $\Omega/n > 1$, both the tidal and the wind torques act to spin down the star, i.e. $\dot{\Omega}\leq 0$. The quasi-co-rotation equilibrium of Eq.~\ref{eq:omsync} would then be $\Omega=n\beta$ with $\beta >1$, but this is forbidden by Eq.~\ref{eqldot} because it would imply $\dot{L} > 0$. The equilibrium state is thus possible only when $\Omega/n < 1$, or $\beta < 1$.

 Equation \ref{lagmeth} provides necessary but not sufficient conditions for the minimization problem. The nature of the stationary point must be investigated using a second partial derivative test to assess whether it is a local minimum, maximum, or saddle point.  If the stationary point of the energy under the constraint of AML is a minimum, then it would be a stable equilibrium point if the constraint were constant in time. Since the AML of the star is time-dependent, a minimum of the energy under this constraint can only be a dynamical equilibrium i.e. a pseudo-stable equilibrium. 
 
\subsection{Stability of the equilibrium}
    Following \citet{Hut1980}, we can compute the Hessian matrix $H$ of the energy using the hypothesis that $\vect{L}(\mathbf{x}) - \vect{f }(\boldsymbol{\Omega}_\star) = 0$ to express the total energy as
\begin{multline}
E =  - G \frac{\subrm{M}{\star}\subrm{M}{p}}{2a} + \frac{1}{2} \subrm{C}{p}(\subrm{\omega}{x}^2 +\subrm{\omega}{y}^2+\subrm{\omega}{z}^2 ) + \frac{1}{2 \subrm{C}{\star}}  \left(\bigl(h \sin i +\subrm{C}{p} \subrm{\omega}{x} \bigr)^2 +  \right. \\ 
\left. \bigl(\subrm{C}{p} \subrm{\omega}{y} \bigr)^2 + \bigl(\subrm{f}{z} - h \cos i - \subrm{C}{p}\subrm{\omega}{z}\bigr)^2\right).
\end{multline}
Then the Hessian at an equilibrium configuration  (i.e. solution of Eq. \ref{lagmeth}) takes the form given in Table~\ref{tab:hess} 
with
\begin{equation}
\alpha=\frac{\subrm{M}{\star} \subrm{M}{p}}{\subrm{M}{\star}+ \subrm{M}{p}} a^2 .
\end{equation}
We note that when $e=0$, $\alpha$ takes the simple form $\alpha=h/n$. 

If the Hessian is positive definite at an equilibrium point, then the energy under the constrain of magnetic braking attains a local minimum at that point. If the Hessian has both positive and negative eigenvalues at an equilibrium point, then it is a saddle point. This is also true even if the stationary point is degenerate.
The eigenvalues of $H$ are the real  solutions $x$ of the equation
\begin{equation}\label{rootsdet}
{\rm det}(H-xI)=0
\end{equation} 
where $I$ represents the unit matrix of dimension $(6,6)$. After some algebra, Eq.~\ref{rootsdet} can be shown to be equivalent to
\begin{eqnarray}
\label{p1p2}
\left[ \left(\frac{G\subrm{M}{\star} \subrm{M}{p}}{4a^3} (\frac{\alpha}{C_\star} + \beta -4 )-x\right)\left(\frac{\subrm{C}{p}} {\subrm{C}{\star}}(\subrm{C}{p} +\subrm{C}{\star})-x\right) \right. \nonumber \\
 \left. -\left(\frac{\subrm{C}{p}} {\subrm{C}{\star}} \frac{G\subrm{M}{\star} \subrm{M}{p}}{2a^2n}\right)^2 \right]  \times \nonumber \\
\left(\frac{G\subrm{M}{\star} \subrm{M}{p}}{a}\beta -x \right) \times \left(\frac{\subrm{C}{p}} {\subrm{C}{\star}}(\subrm{C}{p} +\subrm{C}{\star})-x\right) \times \\
\left[ \left(\frac{G\subrm{M}{\star} \subrm{M}{p}}{a}(\frac{\alpha}{\subrm{C}{\star}}+\beta) -x\right)\left(\frac{\subrm{C}{p}} {\subrm{C}{\star}}(\subrm{C}{p} +\subrm{C}{\star})-x\right) \right. \nonumber \\ 
\left. -\left(\frac{\subrm{C}{p}} {\subrm{C}{\star}}\frac{G\subrm{M}{\star} \subrm{M}{p}}{an} \right)^2 \right] =0. \nonumber
\end{eqnarray}
The first two  solutions come from the two factors on the third line of Eq.~\ref{p1p2}:  
\begin{align}
x_1 &=\frac{\subrm{C}{p}}{\subrm{C}{\star}}(\subrm{C}{p} +\subrm{C}{\star})\label{x1}, \\
x_2 &=\frac{G\subrm{M}{\star} \subrm{M}{p}}{a}\beta\label{x2}.
\end{align}
The other factors  of Eq.~\ref{p1p2} are polynomials of second degree whose discriminants can be shown to always be positive regardless of the value of $\beta$. They yield the following four real roots:
\begin{multline}\label{x3}
x_3 = \frac{1}{2}\left(\frac{G\subrm{M}{\star} \subrm{M}{p}}{4a^3} \Bigl(\frac{\alpha}{C_\star} + \beta -4 \Bigr) +\frac{\subrm{C}{p}}{\subrm{C}{\star}} (\subrm{C}{p} +\subrm{C}{\star}) \right)\\
-\frac{1}{2}\left(\left(\frac{G\subrm{M}{\star} \subrm{M}{p}}{4a^3}(\frac{\alpha}{C_\star} + \beta -4 ) + \frac{\subrm{C}{p}}{\subrm{C}{\star}}(\subrm{C}{p} +\subrm{C}{\star})\right)^2 \right. \\
\left.- \frac{G\subrm{M}{\star} \subrm{M}{p}}{a^3}(\frac{\alpha}{C_\star} + \beta -4 )\frac{\subrm{C}{p}}{\subrm{C}{\star}}(\subrm{C}{p} +\subrm{C}{\star}) +\left(\frac{\subrm{C}{p}}{\subrm{C}{\star}} \frac{G\subrm{M}{\star} \subrm{M}{p}}{na^2}\right)^2 \right)^{1/2},
\end{multline}
\begin{multline}\label{x4}
x_4 = \frac{1}{2}\left(\frac{G\subrm{M}{\star} \subrm{M}{p}}{4a^3} \Bigl(\frac{\alpha}{C_\star} + \beta -4 \Bigr) +\frac{\subrm{C}{p}}{\subrm{C}{\star}} (\subrm{C}{p} +\subrm{C}{\star}) \right)\\
+\frac{1}{2}\left(\left(\frac{G\subrm{M}{\star} \subrm{M}{p}}{4a^3}(\frac{\alpha}{C_\star} + \beta -4 ) + \frac{\subrm{C}{p}}{\subrm{C}{\star}}(\subrm{C}{p} +\subrm{C}{\star})\right)^2 \right. \\
\left.- \frac{G\subrm{M}{\star} \subrm{M}{p}}{a^3}(\frac{\alpha}{C_\star} + \beta -4 )\frac{\subrm{C}{p}}{\subrm{C}{\star}}(\subrm{C}{p} +\subrm{C}{\star}) +\left(\frac{\subrm{C}{p}}{\subrm{C}{\star}} \frac{G\subrm{M}{\star} \subrm{M}{p}}{na^2}\right)^2 \right)^{1/2},
\end{multline}
\begin{multline}\label{x5}
x_5 = \frac{1}{2}\left(\frac{G\subrm{M}{\star} \subrm{M}{p}}{a} \Bigl(\frac{\alpha}{C_\star} + \beta\Bigr) +\frac{\subrm{C}{p}}{\subrm{C}{\star}} (\subrm{C}{p} +\subrm{C}{\star}) \right)\\
-\frac{1}{2}\left(\left(\frac{G\subrm{M}{\star} \subrm{M}{p}}{a}(\frac{\alpha}{C_\star} + \beta) + \frac{\subrm{C}{p}}{\subrm{C}{\star}}(\subrm{C}{p} +\subrm{C}{\star})\right)^2 \right. \\
\left.- 4\frac{G\subrm{M}{\star} \subrm{M}{p}}{a}(\frac{\alpha}{C_\star} + \beta)\frac{\subrm{C}{p}}{\subrm{C}{\star}}(\subrm{C}{p} +\subrm{C}{\star}) +4\left(\frac{\subrm{C}{p}}{\subrm{C}{\star}} \frac{G\subrm{M}{\star} \subrm{M}{p}}{na}\right)^2 \right)^{1/2},
\end{multline}
\begin{multline}\label{x6}
x_6 = \frac{1}{2}\left(\frac{G\subrm{M}{\star} \subrm{M}{p}}{a} \Bigl(\frac{\alpha}{C_\star} + \beta\Bigr) +\frac{\subrm{C}{p}}{\subrm{C}{\star}} (\subrm{C}{p} +\subrm{C}{\star}) \right)\\
+\frac{1}{2}\left(\left(\frac{G\subrm{M}{\star} \subrm{M}{p}}{a}(\frac{\alpha}{C_\star} + \beta) + \frac{\subrm{C}{p}}{\subrm{C}{\star}}(\subrm{C}{p} +\subrm{C}{\star})\right)^2 \right. \\
\left.- 4\frac{G\subrm{M}{\star} \subrm{M}{p}}{a}(\frac{\alpha}{C_\star} + \beta)\frac{\subrm{C}{p}}{\subrm{C}{\star}}(\subrm{C}{p} +\subrm{C}{\star}) +4\left(\frac{\subrm{C}{p}}{\subrm{C}{\star}} \frac{G\subrm{M}{\star} \subrm{M}{p}}{na}\right)^2 \right)^{1/2}.
\end{multline}
From Eq.~\ref{x1}  it is clear that $x_1>0$.  From Eq.~\ref{x2} we have
\begin{equation}
x_2 >0 \Leftrightarrow \beta >0.
\end{equation}
Next, we have from Eq.~\ref{x3}
\begin{equation}
x_3 >0 \Leftrightarrow \beta >4 - \frac{\alpha}{\subrm{C}{p} +\subrm{C}{\star}}.
\end{equation}
It follows from Eq.~\ref{x4} that $x_3>0 \Rightarrow x_4>0$. Finally, Eq.~\ref{x5} yields
\begin{equation}
x_5 >0 \Leftrightarrow \beta >- \frac{\alpha}{\subrm{C}{p} +\subrm{C}{\star}}, 
\end{equation}
and Eq.~\ref{x6} provides the last condition $x_5>0 \Rightarrow x_6>0$. Note that $x_2>0 \Rightarrow x_5>0$. We thus conclude that there are two conditions to be fullfilled to have a pseudo-stable equilibrium :
\begin{equation}
\beta >4 - \frac{\alpha}{\subrm{C}{p} +\subrm{C}{\star}} 
\label{seccondeq}
\end{equation}
and
\begin{equation}
\beta > 0. \label{fircondeq}
\end{equation}
Since at equilibrium $\beta=\Omega/n$, Eq.~\ref{fircondeq} means that the equilibrium is pseudo-stable only for prograde orbits. From hereon, we use the term stable equilibrium to denote what is actually a pseudo-stable equilibrium of this kind. 

 Rewriting Eq.~\ref{seccondeq} at equilibrium, we can show that this point is characterized by
\begin{equation}\label{eq:cond3}
h > ( 4-\beta) (\subrm{C}{p} +\subrm{C}{\star}) n, \\
\end{equation} 
which means that the orbital angular momentum at a stable equilibrium point is greater than $4-n/\Omega$ times the total spin momentum that we would have if the stellar rotation were synchronized with the orbit. 

Our equations reduce to the classical result when magnetic braking is neglected. Indeed, in this case we have in our formulation $\beta=1$ and Eq.~\ref{fircondeq} is always fulfilled. This also means that $n=\Omega$ at equilibrium and the stability criterion is then $h>3  (\subrm{C}{p} +\subrm{C}{\star})n $, as already found by \citet{Hut1980}.

\begin{figure}
   \centering
   \includegraphics[width=\hsize]{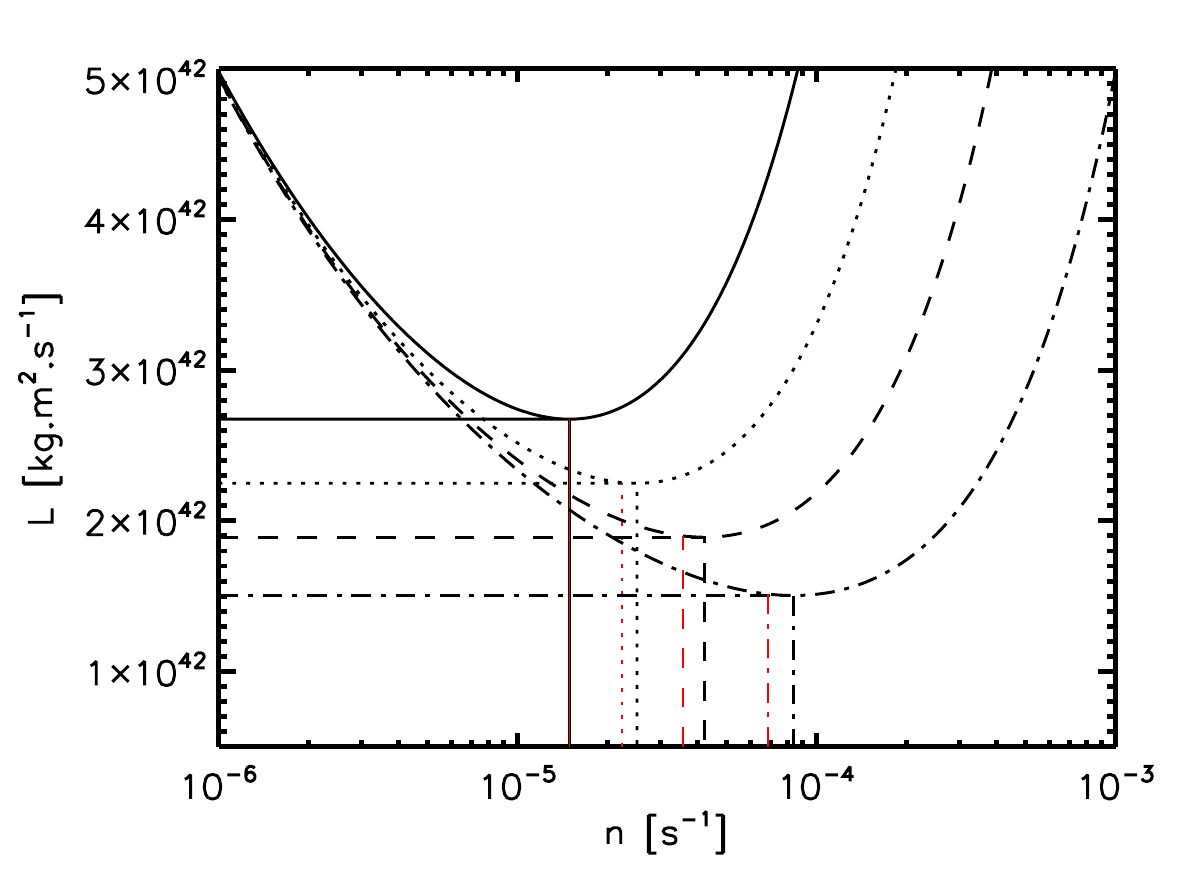}   
      \caption{Quasi-co-rotation curves given by Eq.~\ref{eq:Ltotcorot} for different values of $\beta$, from top to bottom $\beta $=1, 0.5, 0.25, and 0.1. The critical values $L_{\rm c}$ and $n_{\rm c}$ are indicated  by vertical and horizontal black lines whose linestyle corresponds to the different values of $\beta$. The red vertical lines indicate the value of $n_{\rm s}$, again with the linestyle  corresponding to the different values of $\beta$. On each plot, the quasi-co-rotation condition  corresponds to a pseudo-stable equilibrium for $n < n_{\rm s}$ (i.e. on the part of  the curve on the left of the red line $n= n_{\rm s}$), provided that the wind torque dominates the tidal torque. The numerical values correspond to a 1 M$_\odot$ star and a 1 M$_{\rm J}$ planet.}
         \label{HutDiag}
   \end{figure}

Thus we have rigorously demonstrated that the existence of a pseudo-stable equilibrium is possible even when taking magnetic braking into account. The equilibrium is characterized by quasi-co-rotation, which follows from Eq. \ref{ltotgen} to be
\begin{equation}\label{eq:Ltotcorot}
L =   G^{2/3}\frac{M_\star \subrm{M}{p}}{(M_\star + \subrm{M}{p})^{1/3} } n^{-1/3} + ( \beta\subrm{C}{\star} + \subrm{C}{p}) n. 
\end{equation}
 This means that the quasi-co-rotation is possible only when the total angular momentum exceeds a critical value $\subrm{L}{c}$ given by
\begin{equation}\label{eq:Lcrit}
L_{\rm c} =  4 {\left( \frac{G^2}{3^3}\frac{M_\star^3 \subrm{M}{p}^3}{M_\star + \subrm{M}{p} } ( \beta\subrm{C}{\star} + \subrm{C}{p})\right)}^{1/4}. 
\end{equation}
This value depends on $\beta$, which is time dependent. Thus as the system evolves the conditions for the existence of equilibrium also change, but the pseudo-stable equilibrium state can be reached only when $0<\beta <1$. When $L > \subrm{L}{c}$ there are two orbital mean motions consistent with the quasi-co-rotation condition, which are pseudo-stable if the orbital angular momentum $h$ exceeds a value $h_{\rm s} \equiv ( 4-\beta) (\subrm{C}{p} +\subrm{C}{\star}) n$. Contrary to the case where magnetic braking is neglected, this condition does not always mean that one equilibrium is stable while the other is not when $L> \subrm{L}{c}$. Indeed, at $L=\subrm{L}{c} $ the unique mean motion corresponding to quasi-co-rotation is
\begin{equation}
n_{\rm c} =  {\left(  \frac{G^2}{3^3}\frac{M_\star^3 \subrm{M}{p}^3}{M_\star + \subrm{M}{p} } \right)}^{1/4} (\beta \subrm{C}{\star} + \subrm{C}{p})^{-3/4},
\end{equation}
 but the unique mean motion corresponding to quasi-co-rotation and $h=h_{\rm s} $ is
 \begin{equation}
n_{\rm s} =  {\left(  G^2 \frac{M_\star^3 \subrm{M}{p}^3}{M_\star + \subrm{M}{p} } \right)}^{1/4} (4-\beta)^{-3/4} (\subrm{C}{\star} + \subrm{C}{p})^{-3/4}.  
\end{equation}
At $n=n_{\rm s}$ the total angular momentum is 
 \begin{equation}
L_{\rm s} =  {\left(  G^2 \frac{M_\star^3 \subrm{M}{p}^3}{M_\star + \subrm{M}{p} } (4-\beta) (\subrm{C}{\star} + \subrm{C}{p}) \right)}^{1/4} \left( 1 + \frac{\beta \subrm{C}{\star} + \subrm{C}{p}}{ (4-\beta) (\subrm{C}{\star} + \subrm{C}{p})} \right).
\end{equation}
Thus
\begin{equation}
\frac{n_{\rm s}}{n_{\rm c}} = 3^{3/4} \left(\frac{\beta \subrm{C}{\star} + \subrm{C}{p}}{ (4-\beta) (\subrm{C}{\star} + \subrm{C}{p})} \right)^{3/4}
\end{equation}
and 
\begin{equation}
\frac{L_{\rm s}}{L_{\rm c}} = \frac{3^{3/4}}{4}  {\left( \frac{(4-\beta) (\subrm{C}{\star} + \subrm{C}{p} )}{\beta \subrm{C}{\star} + \subrm{C}{p}} \right)}^{1/4} \left( 1 + \frac{\beta \subrm{C}{\star} + \subrm{C}{p}}{ (4-\beta) (\subrm{C}{\star} + \subrm{C}{p})} \right).
\end{equation}
Only when $\beta=1$ do we have $n_{\rm c}= n_{\rm s}$, $L_{\rm c}= L_{\rm s}$ and one stable and one unstable equilibrium state, as in Hut's theory. Since we have already established that $0 < \beta < 1$ at a pseudo-stable equilibrium point, there is one stable and one unstable equilibrium if $L > L_{\rm s}$, since we then have $n_{\rm s} < n_{\rm c}$ and $L_{\rm s} > L_{\rm c}$. The stable equilibrium has $n<n_{\rm s}$ because it requires $h> h_{\rm s}$, and $h$ increases with decreasing $n$, as can be seen by considering its expression in the first term on the right-hand side of Eq.~\ref{eq:Ltotcorot}.  This can be seen in Fig.~\ref{HutDiag}, where the quasi-co-rotation curves are given for different values of $\beta$, and the corresponding values of $L_{\rm c}$, $n_{\rm c}$, and $n_{\rm s}$ are shown. When $L_{\rm c}<L<L_{\rm s}$, the two quasi-synchronous states corresponding to $L$ are unstable because $n > n_{\rm s}$ and $h < h_{\rm s}$.

In conclusion, if $L>\subrm{L}{s}$, there are two possible equilibrium states, one that is stable, and the other one unstable. This generalizes the criteria for stability previously defined by \citet{Darwin1879}. Since $\beta$ evolves in time when we account for the AML, a system that is at some time Darwin stable might become unstable during its evolution. But if the system can keep $0 <\beta <1$ and $n < n_{\rm s}$, it will evolve along a series of stable states with $\Omega=\beta n$. As long as the wind torque dominates the tidal torque, i.e. $\dot{\Omega} <0$, the in-fall of the planet into the star is delayed. On the other hand, if $L<\subrm{L}{c}$  or $L_{\rm c} \leq L \leq L_{\rm s}$, there is no pseudo-stable equilibrium possible, and the system can be considered Darwin unstable.  This formally establishes the conditions for quasi-equilibrium when accounting for AML, so in this way, the tidal evolution of exoplanetary systems can indeed be studied in terms of Darwin stability. 

\section{Pseudo-stability of hot Jupiters}\label{stabHJ}
For the reasons explained in Sec.\ref{effmag}, we only consider the case of circular and aligned systems. Let us consider the critical angular momentum $\subrm{L}{\rm c}$ in the absence of magnetic braking, i.e. when $\beta=1$. We denote it as 
\begin{equation}\label{eq:Lcrit0}
L_{\rm c_0} =  4 {\left[  \frac{G^2}{3^3}\frac{M_\star^3 \subrm{M}{p}^3}{M_\star + \subrm{M}{p} } ( \subrm{C}{\star} + \subrm{C}{p})\right]}^{1/4}.  
\end{equation}
At $L=\subrm{L}{c_0} $, the unique mean motion corresponding to co-rotation, in the absence of magnetic braking, is
\begin{equation}\label{nc0}
n_{\rm c_0} =  {\left(  \frac{G^2}{3^3}\frac{M_\star^3 \subrm{M}{p}^3}{M_\star + \subrm{M}{p} } \right)}^{1/4} (\subrm{C}{\star} + \subrm{C}{p})^{-3/4}.
\end{equation}
 The values of $\subrm{L}{c_0}$ and $\subrm{n}{c_0}$ only depend on the masses and radii of the star and the planet, and there is no need to know the actual value of $\beta$ or even the form of $f(\Omega)$ to compute them. Using these notations, the total angular momentum of a circular and aligned system in units of the critical angular momentum can be written as 
\begin{equation}\label{eqLtotLc}
\frac{L}{\subrm{L}{c_0}} = \frac{1}{4}\left(3 \left(\frac{n}{\subrm{n}{c_0}}\right)^{-1/3} +\ \frac{\subrm{C}{\star}}{\subrm{C}{\star}+\subrm{C}{p}} \frac{\Omega}{\subrm{n}{c_0}}\ +\ \frac{\subrm{C}{p}}{\subrm{C}{\star}+\subrm{C}{p}} \frac{\omega}{\subrm{n}{c_0}}\right).
\end{equation}
 The first term on the right-hand side of Eq. (\ref{eqLtotLc}) corresponds to the contribution of the orbital angular momentum $h$, the second and third terms to the stellar and planetary rotational momenta $\subrm{L}{\star}$ and $\subrm{L}{p}$, respectively. We  notice the simple relationship between the critical values with and without AML:
\begin{figure}
	\centering
        \includegraphics[width=\hsize]{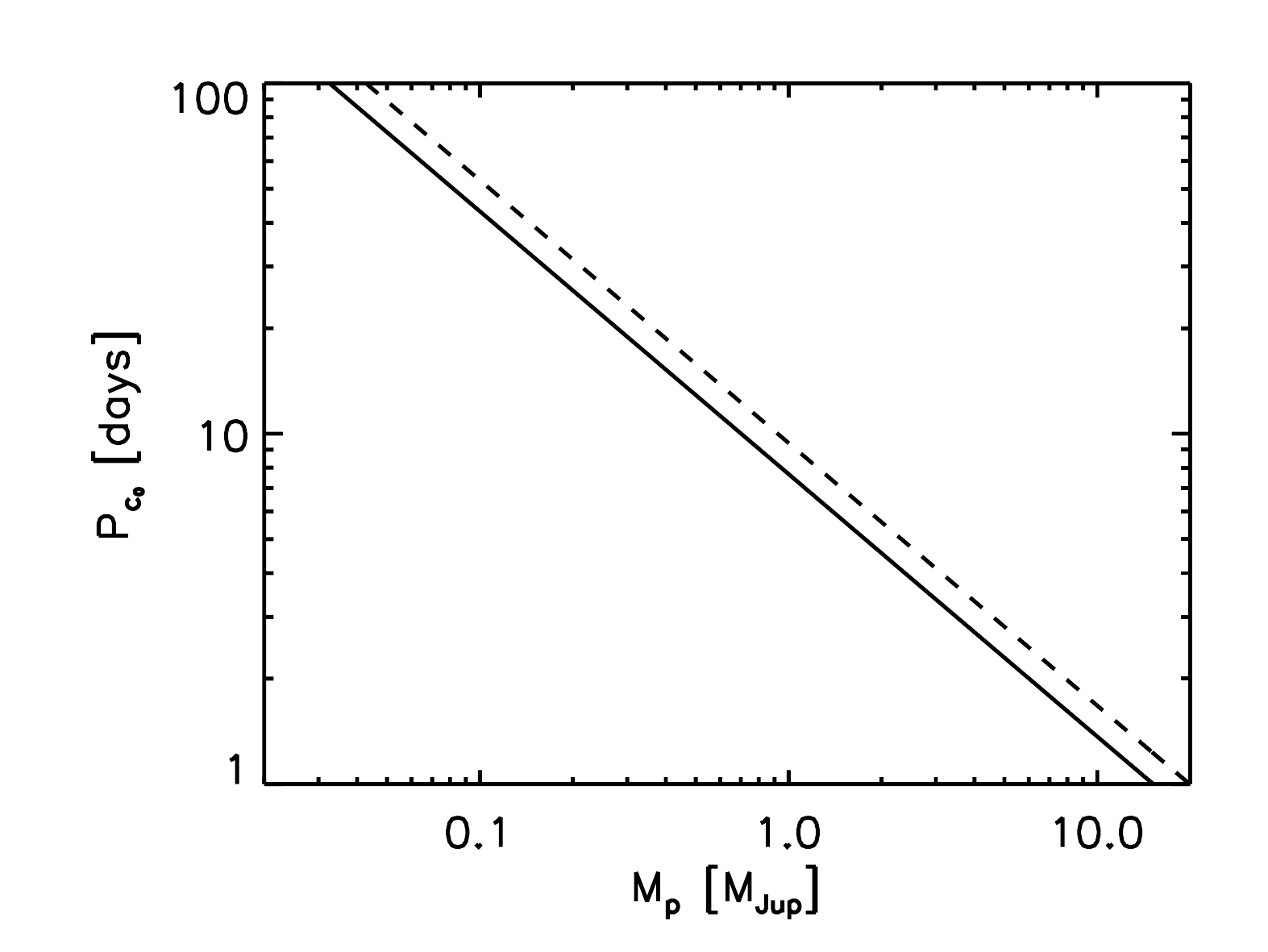}
     	 \caption{Critical orbital period as a function of planetary mass for  a stellar mass and radius corresponding to a G0 (solid) or F0 (dashed) main-sequence star. If $L < \subrm{L}{c}$, systems with period shorter than the critical one are Darwin unstable.}
         \label{pcritmplan}
\end{figure}
\begin{figure*}
	\centering
        \includegraphics[width=0.8\hsize]{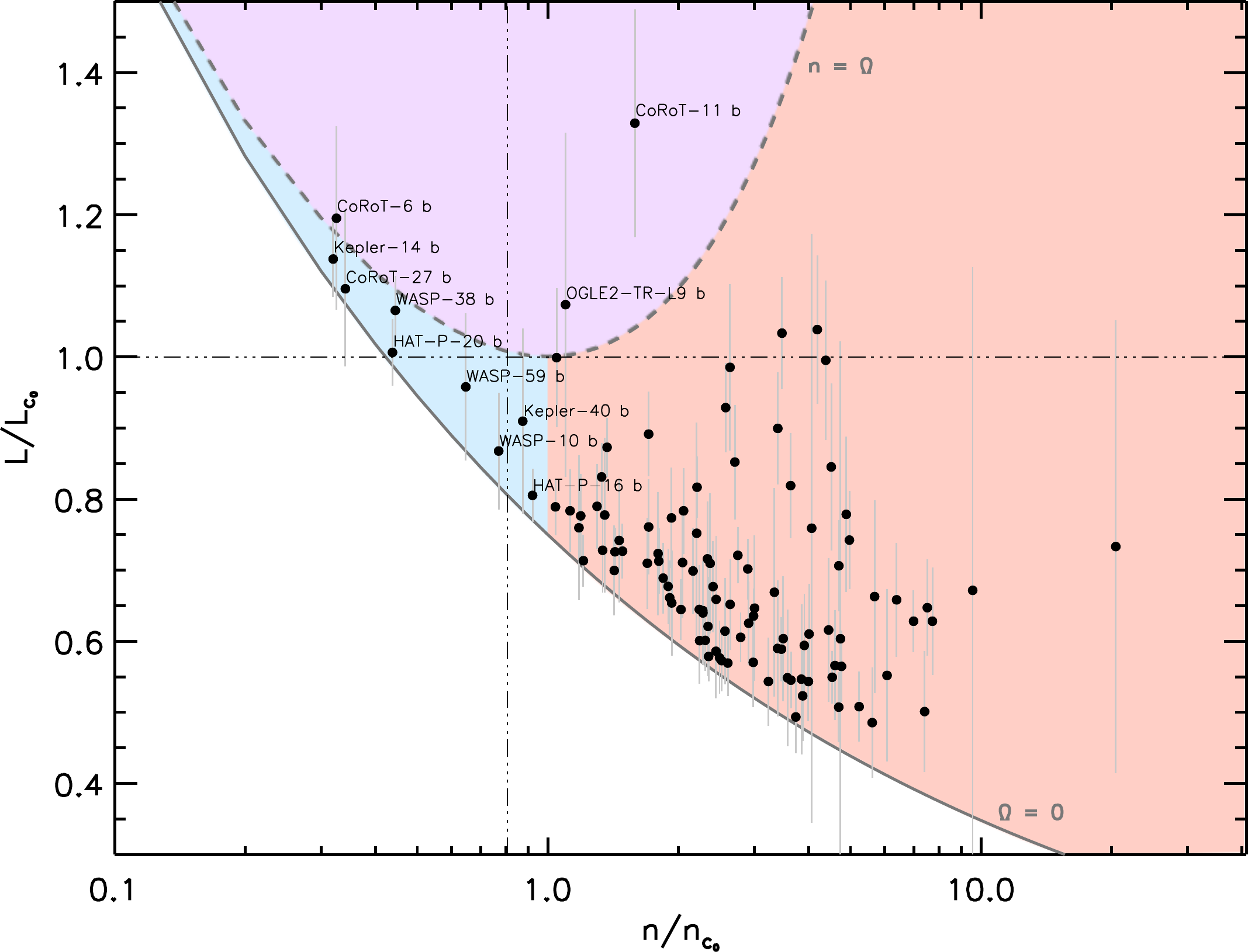}
       	\caption{Darwin diagram showing the 109 transiting systems known to date with negligible obliquity and eccentricity. The total angular momentum in units of the critical angular momentum for a conservative system is plotted vs. the observed mean motion of the orbit of the planet in units of the critical mean motion. The solid line is, for a given orbit, the contribution of the orbital angular momentum to the total momentum, while the dashed line is the locus of spin-orbit synchronization. Thin dashed-three-dotted lines indicate the values $L= \subrm{L}{c_0}$ and $n=(3/4)^{3/4}\subrm{n}{c_0}$, respectively. The red domain corresponds to systems that cannot be evolving towards a stable state, whatever the efficiency of tides or AML. The purple and blue domains contain systems that could be evolving towards their stable state, while some of the systems in the blue domain could already be in their asymptotically stable orbit.}
         \label{darwindiag}
\end{figure*}
\begin{align}
\frac{\subrm{L}{c}}{\subrm{L}{c_0}} &= \left(\frac{\beta  C_\star + \subrm{C}{p}}{C_\star +\subrm{C}{p}}\right)^{1/4} \qquad (\beta>0),\\
\frac{\subrm{n}{c}}{\subrm{n}{c_0}} &= \left(\frac{\beta  C_\star + \subrm{C}{p}}{C_\star +\subrm{C}{p}}\right)^{-3/4}\qquad (\beta>0),\\
\frac{\subrm{L}{s}}{\subrm{L}{c_0}} &= \frac{3^{3/4}}{4}( 4-\beta)^{1/4} \left(1 + \frac{\beta  C_\star + \subrm{C}{p}}{C_\star +\subrm{C}{p}}\right)^{1/4} \qquad (\beta<4),\\
\frac{\subrm{n}{s}}{\subrm{n}{c_0}} &=  \left(\frac{3}{4-\beta}\right)^{3/4} \qquad (\beta < 4).
\end{align}
 Considering that for typical  Jupiter-sized planets $\subrm{C}{p}< 10^{-4} C_\star$, we can neglect $C_{\rm p}$ and write
\begin{align}
\frac{L}{\subrm{L}{c_0}} &\approx \frac{1}{4} \left(3 \left(\frac{n}{\subrm{n}{c_0}}\right)^{-1/3} +\ \frac{\Omega}{n}\frac{n}{\subrm{n}{c_0}}\ \right),\label{eqLtotLc2}\\
\frac{\subrm{L}{c}}{\subrm{L}{c_0}} &\approx \beta^{1/4} \qquad (\beta>0),\\
\frac{\subrm{n}{c}}{\subrm{n}{c_0}} &\approx \beta^{-3/4}\qquad (\beta>0),\\
\frac{\subrm{L}{s}}{\subrm{L}{c_0}} &\approx \left(\frac{3}{4-\beta}\right)^{3/4} \qquad (\beta < 4).
\end{align}
A given system can have a \rm{pseudo-}stable equilibrium if $0<\beta<1$, which implies $\subrm{L}{c} < \subrm{L}{s} < \subrm{L}{c_0}$ and $\subrm{n}{s} < \subrm{n}{c_0} < \subrm{n}{c}$. In Fig.~\ref{pcritmplan}, we illustrate the values of the critical orbital period $\subrm{P}{c_0}$ corresponding to $\subrm{n}{c_0}$ for a range of planetary mass and stellar parameters, neglecting $C_{\rm p}$.

Using the dimensionless form given by Eq.~\ref{eqLtotLc2}, we can plot all the known systems in a Darwin diagram  and see how they relate to the quasi-synchronous (or pseudo-equilibrium) state.  Considering the Exoplanet Orbit Database\footnote{as of May 2014, see http://exoplanets.org}, we selected a sample of transiting planets that orbit a single F-, G-, or K-type star, with known planetary mass and stellar rotation rate, and for which no additional companion has been detected. Furthermore, we restrict our study to the subsample of systems with negligible eccentricity and projected obliquity $\lambda$ ($e<0.1$ and $|\lambda| < 30^\circ$). This results in a sample of 109 systems whose parameters are listed in Table \ref{tab:param}. 

We plot in  Fig. \ref{darwindiag} the total angular momentum as a function of the observed mean motion of the orbit in units of $\subrm{L}{c_0}$ and $\subrm{n}{c_0}$, respectively. We assumed that the rotation of the planet has already reached synchronization with the orbit, but, as said previously, the spin of the planet is negligible. To compute the moments of inertia we used, for the star, the gyration radius given by the models of \citet{Claret1995} as a function of mass and $\subrm{T}{eff}$, while, for the planet, we considered a polytropic model of index 1. To estimate the rotation rate of the star, we computed it from the measured projected rotational velocity $v \sin i$ and stellar radius. Considering that we have selected transiting systems with negligible obliquity, it is reasonable to assume that $\sin i \approx 1$, but strictly speaking we obtain a lower limit on $L$. 

Since Darwin stable systems must have $ L > \subrm{L}{s}$, but $\subrm{L}{s} < \subrm{L}{c_0}$, comparing the value of the current total angular momentum $L$ to the critical value $\subrm{L}{c_0}$ does not necessarily allow  their stability to be inferred. Nevertheless, we can reach a conclusion for some of them as indicated in Fig.~\ref{darwindiag}. First, for the few systems that currently have $L>\subrm{L}{c_0}$ and $\Omega>n$ (highlighted in purple in Fig.~\ref{darwindiag}), the star is necessarily spun down both by the tides and the wind. Those systems are not currently in a stable state but could  evolve towards it. Second, the systems that have $\Omega<n$ could have $0<\beta<1$ and $ L > \subrm{L}{s}$, so that they could have a possible stable state. However the stable state in this case would imply $n <\subrm{n}{s} < \subrm{n}{c_0}$. Since in this part of the diagram, the tides can only bring the planet closer to the star, the stable state is impossible to reach for systems that have $n>\subrm{n}{c_0}$ (highlighted in red in Fig.\ref{darwindiag}). Regardless of the value of $\beta$, they will eventually plunge into their star, but those that have $0<\beta<1$ could first evolve towards their unstable equilibrium state. When $\beta>1$, there is no equilibrium possible, and the planet falls directly into the star. Third, the systems that have $n< \subrm{n}{c_0}$ and $\Omega<n$ (highlighted in blue in Fig.\ref{darwindiag}) are potentially Darwin stable. They can evolve towards their pseudo-stable state if $0<\beta<1$ and $L>\subrm{L}{s}$. Since $\subrm{n}{s}/\subrm{n}{c} < (3/4)^{3/4}$ when $\beta<0$, the systems that have  $n/\subrm{n}{c_0} < (3/4)^{3/4}$ could already be in their pseudo-stable state if $0<\beta<1$.

As previously noted by \citet{Matsumura2010}, we see at once in Fig. \ref{darwindiag} that most of the systems of our sample are indeed Darwin unstable. But accounting for the balance between the magnetic braking and the tidal torque creates a new possibility for equilibrium. \citet{Matsumura2010} find that all planetary systems known at the time were unstable except CoRoT-3, CoRoT-6, HD~80606, and WASP-7. Our results agree for CoRoT-6, but we have rejected CoRoT-3 and WASP-7 from our sample because they have a measured $|\lambda| > 30^{\circ}$ and  HD~80606 because the star has a stellar companion. We find in addition that our updated list contains 18 more systems that fulfil the condition $L \gtrsim \subrm{L}{c_0}$ within the error bars, but  for most of them it is only marginally significant, and only four systems  have $L > \subrm{L}{c_0}$ significant at more than $1 \sigma$. They are CoRoT-11, CoRoT-6, Kepler-14, and WASP-38. Among these, two have $n < \Omega$, therefore, they are migrating outwards and are thus currently Darwin pseudo-stable. Kepler-14 and WASP-38, on the other hand, have $n> \Omega$, and with most of their angular momentum in the form of orbital momentum ($h=26 \pm 6 (\subrm{L}{\star}+\subrm{L}{p}$) and $h=12 \pm 1 (\subrm{L}{\star}+\subrm{L}{p})$ respectively), they also have $n/\subrm{n}{c_0} < (3/4)^{3/4}$. They could be in the pseudo-stable state. If we also consider the marginally significant systems with $L > \subrm{L}{c_0}$, we have 14 other systems. Among these, ten have $n>\subrm{n}{c_0}$ and cannot evolve toward a stable equilibrium. The other four are CoRoT-27, HAT-P-20, Kepler-40, and  WASP-59, and they could all be in their pseudo-stable state. In addition, there are two more systems that could be Darwin stable even if they have $L<\subrm{L}{c_0}$, namely HAT-P-16 and WASP-10.
 
 Finally,  we note that HAT-P-20 might be mistakenly included in our sample because a stellar neighbour has been detected, although the gravitational bound to the primary remains to be confirmed \citep{Bakos2011}. We also note that the parameters inferred from the first analysis of CoRoT-11 suggested a circular orbit \citep{Gandolfi2010}, and this is the value we have adopted. However, a recent re-analysis of CoRoT light curves based on a Bayesian model selection \citep{Parviainen2013} claims to have detected a statistically significant secondary eclipse, and inferred an eccentricity of the orbit $e=0.35\pm 0.03$ from its phase. They have neither performed a detailed light curve modelling nor considered consistency with radial velocity data, but if this value is confirmed, this would potentially change the derived parameters of the planet. Not only would this system be discarded from our sample, but a detailed study would be impossible until a consistent fit of the light curve and radial velocity data is available.
 
Even without detailed knowledge of the AML law or tidal dissipation mechanisms, the use of Darwin diagrams allow the assessment of the state of tidal evolution and its likely outcome. For circular and aligned systems, the evolution of the orbital elements only depends on the initial distribution of the angular momentum between the spin of the star and the orbit. Darwin diagrams can be used to infer the past evolution of the systems, as we show in the next section.
\section{Evolution in the Darwin diagram}\label{SecEvol}
The temporal evolution of the orbital parameters depends on the efficiency of tidal dissipation and magnetic braking, which are currently theoretical challenges as reviewed in Sects.~\ref{tidtheo} and ~\ref{magbrak}. If their actual values are not well known, we expect, however, to observe a qualitative difference between F- and later-type stars, because both tidal dissipation and magnetic braking are related to the extension of their convective zone. We have rigorously demonstrated that $\Omega=n$ is not an equilibrium state when including magnetic braking.  The  conditions for the existence of a pseudo-stable orbit are time dependent and, if  the evolution could proceed indefinitely, the continuous loss of angular momentum from the star would eventually bring any systems to an energy state where no equilibrium is possible. Nevertheless, the existence of a dynamical equilibrium state is possible even when $L<\subrm{L}{c_0}$. In this case, the orbit does not necessarily shrink exponentially, but can be first brought towards the quasi-equilibrium state, which is a time-dependent function of the magnetic braking law. The pseudo-equilibrium state requires $\dot{\Omega}<0$, meaning that angular momentum loss via stellar wind must compensate for the angular momentum gain from the orbit as the planet attempts to spin up the star. 

\begin{figure}[h]
   \centering
   \includegraphics[width=1.0\hsize]{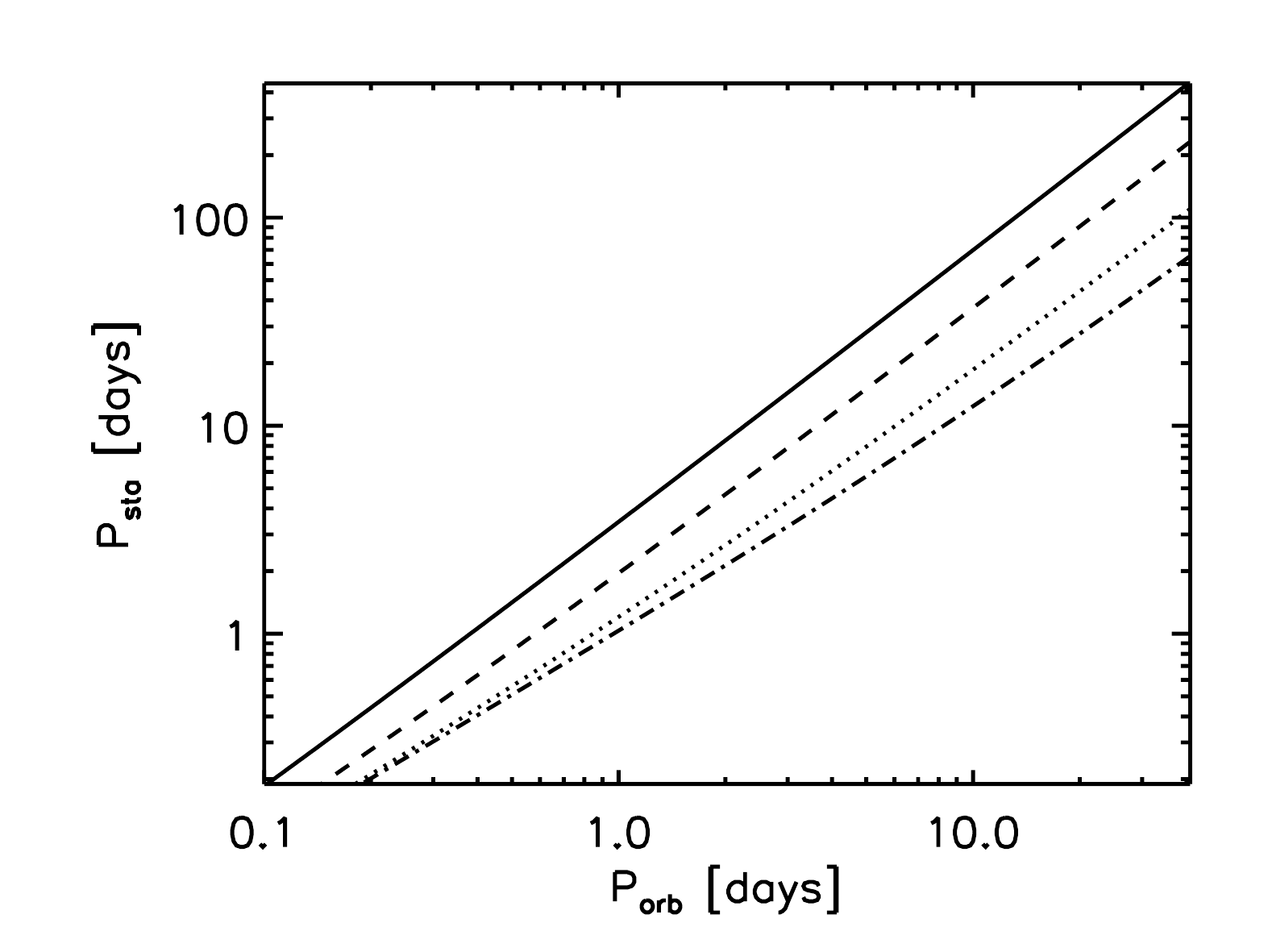}
   \caption{Stellar rotation period corresponding to torque balance as a function of the orbital period in days. The solid line is for Model \#1, the dotted line for  Model \#2, the dashed line for  Model \#3, and the dash-dotted line for  Model \#4. }
   \label{synchstat}
\end{figure}

To find the location of the equilibrium, we need to assume some form of tidal dissipation and magnetic braking. We use a formulation based on \citet{Barker2009}, obtained in the framework of the equilibrium tide assuming a constant $Q\rq{}$. Adopting a constant $Q\rq{}$ implies that the time lag between the maximum of the tidal potential and the tidal bulge in each body scales with the orbital period and that the relevant tidal frequency is the orbital frequency. This may not give identical numerical factors in the resulting equations to other formulations of tidal friction \citep{Goldreich1966, Zahn1977, Hut1981, Matsumura2008}. Given our uncertainties on the value of $Q\rq{}$ and its dependence on the tidal frequency, we feel this is the most practical way to study the general effects of tidal friction. We use a Skumanich-type law for magnetic braking with a torque of magnitude $\subrm{\Gamma}{mb}=- \subrm{\alpha}{mb} C_\star \Omega^3$, where the value of $\subrm{\alpha}{mb}$ is estimated from observed rotational velocities of stars in clusters of different ages. Neglecting  tides in the planet and the planetary spin, the following set of dimensionless equations can be used to describe the temporal evolution of the stellar spin frequency and the orbital mean motion, 
\begin{align}
\frac{{\rm d}\tilde{\Omega}}{{\rm d}\tilde{t}} &= \tilde{n}^4 \left(1-\frac{\tilde{\Omega}}{\tilde{n}}\right) - A \tilde{\Omega}^3,\label{omegadot}\\
\frac{{\rm d}\tilde{n}}{{\rm d}\tilde{t}} &= 3 \tilde{n}^{16/3} \left(1-\frac{\tilde{\Omega}}{\tilde{n}}\right),\label{ndot}
\end{align}
where $\tilde{\Omega}$ and $\tilde{n}$ are dimensionless variables that are related to the ones previously defined by the following relationships: 
\begin{align}
\tilde{n} = \frac{n}{\subrm{n}{c_0}} 3^{-3/4}, & \qquad
\tilde{\Omega} = \frac{\Omega}{\subrm{n}{c_0}} 3^{-3/4},
\end{align}
and $A$ is a non-dimensional constant defined as
\begin{equation}\label{eq:a}
A=\frac{2}{3^{9/4}}\subrm{\alpha}{mb} Q^\prime\subrm{n}{c_0}\frac{\subrm{M}{\star}}{\subrm{M}{p}} \subrm{r}{g}^{5}  \left(\frac{\subrm{M}{p}}{\subrm{M}{\star}+\subrm{M}{p}}\right)^{-5/2} .
\end{equation}
The stationary state, i.e when the torque exerted on the star by the wind is balanced by the tidal torque, is equivalent to $\dot{\Omega} =0$. According to Eq.~\ref{omegadot}, this means 
\begin{equation}\label{cubic}
\tilde{\Omega}^3 + \frac{\tilde{n}^3}{A} \tilde{\Omega} -  \frac{\tilde{n}^4}{A} = 0. 
\end{equation}
The discriminant of this cubic equation in $\tilde{\Omega}$ is always negative when $\tilde{n}>0$, thus for each positive $\tilde{n}$ there is one real value $\tilde{\Omega}_{\rm sta}$ corresponding to the torque balance. Using Cardano\rq{}s method, the real root of Eq.\ref{cubic} can be written as
\begin{equation}\label{eqstat}
\tilde{\Omega}_{\rm sta} = \tilde{n} \, \sqrt[3]{\frac{\tilde{n}}{2A}} \left(\sqrt[3]{1+\sqrt{1+\frac{4\tilde{n}}{27A}}} +\sqrt[3]{1-\sqrt{1+\frac{4\tilde{n}}{27A}}  }\right). 
\end{equation}
We can then use this expression in Eq.\ref{eqLtotLc2} and find the locus of total angular momentum yielding $\dot{\Omega} =0$ in a Darwin diagram. The value of $A$ depends on the masses of both planet and star, but also on the uncertain parameters $Q^\prime$ and $\subrm{\alpha}{mb}$. For illustration purposes,  we computed the stationary locus and the evolution of $\Omega$ and $n$ using Eqs.~\ref{omegadot} and \ref{ndot} for four systems: two with a solar-like host and a planet of 1 and 10 $\subrm{M}{J}$, respectively (Models 1 and 2), and the other two for the same planetary masses, but with an F-type star (Models 3 and 4). Following \citet{Barker2009} and \citet{Dobbs-Dixon2004}, we take $\subrm{\alpha}{mb} = 1.5 \times 10^{-14} \gamma$~yr where $\gamma=1.0$  for G stars and $\gamma = 0.1$ for F stars. To better appreciate the effect of magnetic braking, we use the same value of $Q^\prime= 10^7$ for all the models \citep{Jackson2009}. The parameters and the corresponding values of $A$ for the four models are given in Table~\ref{tab:a}. 
\begin{table}
	\caption{Model parameters} 
	\label{tab:a} 
	\centering
	\begin{tabular}{l c c c c c}
	\hline\hline
	Model & $M_\star$ (M$_\odot$) & $\subrm{M}{p}$ (M$_{\rm J}$) & $Q^\prime$ & $\gamma$ & A\\
	\hline
	\#1& 1 &1& $10^7$ &1& 62 \\
	\#2& 1 &10& $10^7$ &1& 0.1 \\
	\#3& 1.4 &1& $10^7$ &0.1& 12 \\
	\#4& 1.4 &10& $10^7$ &0.1& 0.02 \\
	\hline
	\end{tabular}
\end{table}
In the Darwin diagram, the locus of the stationary state defines for each orbital frequency a unique stellar rotational velocity. It is illustrated in Fig.~\ref{synchstat} by displaying the corresponding stellar rotation period as a function of the orbital period.
For the more massive planet, the stationary state is very close to  synchronization  for a wide range of the orbital mean motion, regardless of the mass of the star, whereas the frequency ratio takes a greater value for the Jupiter-size planet, and is more dependent on the mass of the star. 

\begin{figure*}
   \centering
   \includegraphics[width=\hsize]{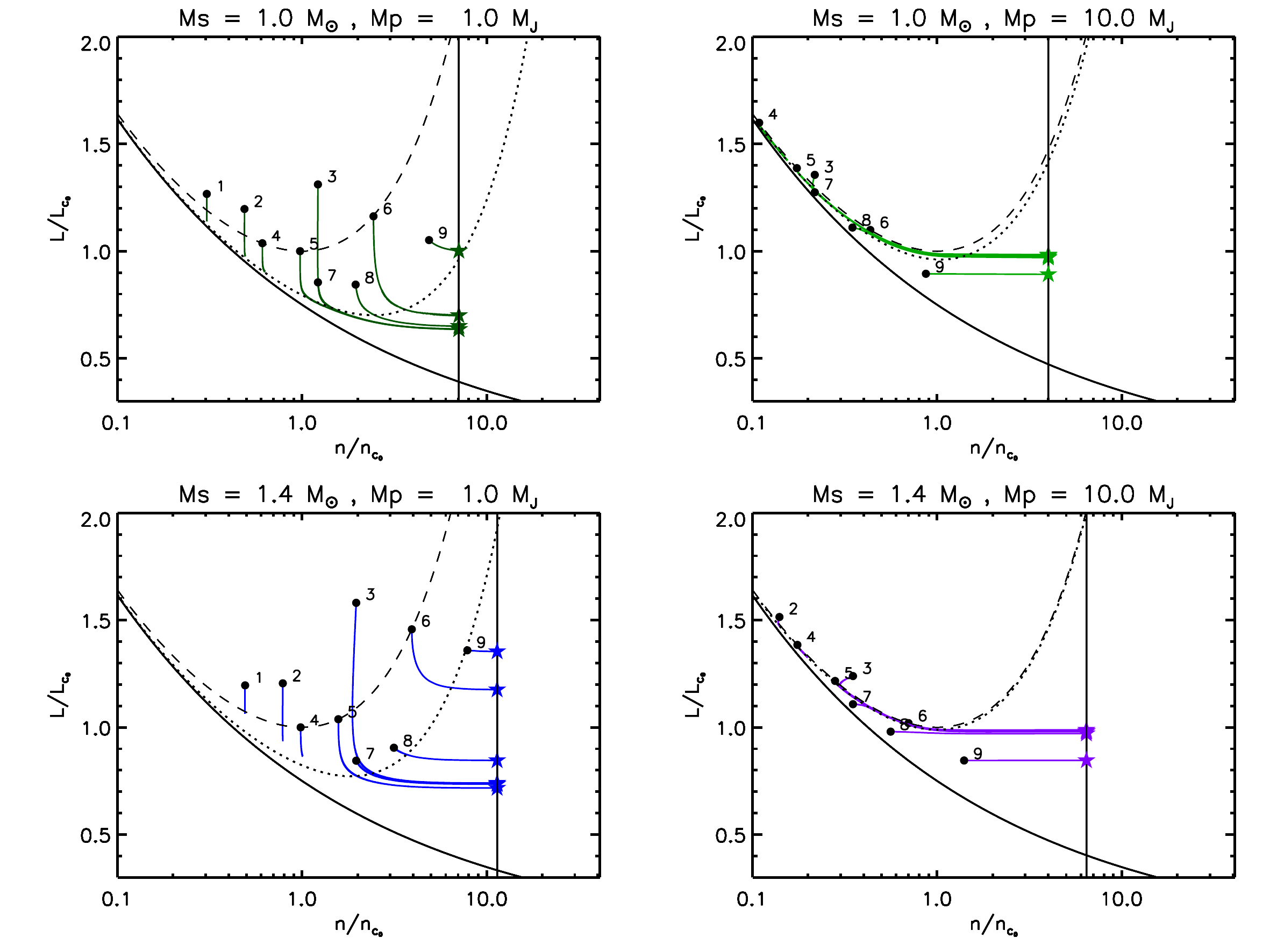}   
   \caption{Darwin diagram showing the evolution of the angular momentum of planetary systems under the action of both tidal dissipation and magnetic braking. The evolutionary tracks are shown in different colours for different combinations of stellar and planetary mass. Models include a solar-like star (top) or an F-type star (bottom) and a Jupiter mass (left) or 10 times the mass of Jupiter (right)  planet. The black dashed line is the locus of synchronization $\Omega=n$, while the dotted line is the locus of balance between the rate of AML due to the wind and the rate of AM transfer with the orbit ($\dot{\Omega}=0$). The solid black vertical line gives the Roche limit where the computation stops. Different initial conditions are indicated by numbers corresponding to their sets as listed in Table~\ref{condint}.  The end of the evolutionary track is symbolized by a star for the planets that reach the Roche limit in less than 13~Gyr.}
     \label{darwindiagFG}
\end{figure*}

To understand how and when a system can reach the locus of stationary rotation rate, we computed for each model several evolutionary paths  characterized by different initial stellar and orbital periods, using stellar periods of 8, 5, or 2~days typical of young stars, and took the initial orbital period as half, equal, or twice the initial stellar rotation period. The different sets of initial conditions are listed in Table~\ref{condint}. We do not regard all the initial conditions considered  here as equally probable, but only assume them for illustration purposes.
\begin{table}
	\caption{Initial parameters  for the different paths shown in Fig.~\ref{darwindiagFG}.}
	\label{condint} 
	\centering
	\begin{tabular}{l c c}
	\hline
	\hline
	Set & $\subrm{P}{rot}$ (days) & $\subrm{P}{orb}$ (days)  \\
	\hline
	1& 8 & 16\\
	2& 5 & 10\\
	3& 2 & 4\\
	4& 8 & 8\\
	5& 5 & 5\\
	6& 2 & 2 \\
	7& 8 & 4 \\
	8& 5 & 2.5\\
	9& 2 & 1 \\
	\hline
	\end{tabular}
\end{table}
We let the system evolve for 13~Gyr or until the planet reaches the Roche limit $\subrm{a}{R}$ defined as
\begin{equation}
\subrm{a}{R} = 2.422\subrm{R}{p}\left(\frac{\subrm{M}{\star}}{\subrm{M}{p}}\right)^{1/3},
\end{equation}
 taking $\subrm{R}{p} = 1.3 \subrm{R}{J}$ for all masses. The results are presented in Fig.~\ref{darwindiagFG}. When the planet reaches the Roche limit before 13 Gyr, we indicate the end of the track by a star. An evolution lasting 13 Gyr is remarkably longer than the main-sequence lifetime of a G-  or a F-type star, but we assume this as a conservative value considering that the tidal dissipation efficiency could be stronger than considered here, which would result in faster evolution.
The global features of tidal evolution under the constraint of AML  and the connection with the pseudo-equilibrium we have found in Sect.~\ref{stabGen} can be understood as follows. When the stellar rotation rate is greater than the orbital frequency $\Omega >n$, the tides and the wind act to spin down the star. In Fig.~\ref{darwindiagFG} this corresponds to the domain above the dashed curve where the orbit is slower than the stellar spin. The first stages of Sets 1, 2, and 3 fall in this part of the diagram. There, the value of the parameter $\beta$ at the pseudo-equilibrium is bounded by $0< \beta < 1$,  as can be deduced from Eq.~(\ref{eqldot}) because $dL/dt < 0$ and $d\Omega/dt <0$. If the wind torque is much greater in magnitude than the tidal torque exerted on the star, (e.g. G-type star  and long-period low-mass planet), the pseudo-equilibrium is characterized by $\beta \rightarrow 0$ because the total angular momentum loss in Eq.~(\ref{eqldot}) corresponds to the decrease in the stellar angular velocity. In this case, the star spins down faster than the orbital adjustment of the planet and its orbital migration is reduced, leading to an almost vertical evolution in the diagram. This can be seen on Tracks 1, 2, and 3, above the dashed line, in the top left-hand panel of Fig.~\ref{darwindiagFG}. If the wind torque is smaller in magnitude than the tidal torque (e.g. F-type star and short-period massive planet), the total angular momentum loss rate decreases and $\beta \rightarrow 1^{-}$.  The planet can migrate outwards toward the pseudo-stable state characterized by $\Omega=\beta n \lesssim n$. This can be seen on Track 3, above the dashed line, in the bottom right-hand panel of Fig.~\ref{darwindiagFG}.

In both cases,  the successive stage of the evolution is characterized by  the spinning down of the star. If the system reaches the locus of synchronization $\Omega=n$, the tidal torque vanishes, and the corresponding $\beta$ at pseudo-equilibrium tends to zero. The evolution towards the minimum of energy proceeds, $\Omega$ becomes smaller than $n$, the orbit is now faster than the stellar spin, and the tidal torque and wind torque have opposite signs. As long as the magnitude of the wind torque is greater than the magnitude of the tidal torque, $\dot{\Omega} <0$, the star keeps braking down, but the magnitude of the wind torque also decreases. In Fig. ~\ref{darwindiagFG} this corresponds to the part of the evolution taking place between the dashed line and the dotted line. It can be described as follows.

If the increase in the tidal torque and the decrease in the magnitude of the wind torque are slow enough that the former remains small compared to the latter (e.g. a long-period planet), then at pseudo-equilibrium $\beta \rightarrow 0^{+}$ as $\dot{L} \rightarrow 0^{-}$ and $\dot{\Omega} \rightarrow 0^{-}$. The system  can evolve towards the pseudo-equilibrium characterized by $\Omega/n=\beta$. As long as there is more orbital angular momentum than $4-\beta$ times the spin momentum that we would have if the stellar rotation were synchronized with the orbit (e.g long-period high-mass planet), the pseudo-equilibrium state is pseudo-stable. In fact, a succession of stable states will be reached, evolving asymptotically towards $\dot{\Omega} \rightarrow 0$, $\dot{L} \rightarrow 0$ and $\beta \rightarrow 0$. This is what happens, for example, along Track 4, in the top right-hand panel of Fig.~\ref{darwindiagFG}. The system remains in a pseudo-stable state for an extended period of time. This results in slow orbital migration and total angular momentum loss over the 13 Gyr of the simulation. If there is not enough orbital angular momentum to ensure the stability of the equilibrium, the system can only evolve towards an unstable pseudo-equilibrium state when $L>\subrm{L}{c}$ and the tidal torque increases. This is what we see for Track 7, in the top right-hand panel of Fig.~\ref{darwindiagFG}. The pseudo-stable state cannot be maintained long enough to prevent the in-fall of the planet within the time span of the simulation.

In any case, no equilibrium point can be indefinitely stable under the constraint of angular momentum loss. As the rising tidal torque becomes larger than the falling wind torque, eventually $\beta \rightarrow -\infty$ as $\dot{\Omega} \rightarrow 0^{-}$, but $\dot{L}$ does not vanish (cf. Eq.~\ref{eqldot}). First, the rising tidal torque approaches the decreasing wind torque and the evolution proceeds at almost constant stellar rotation frequency, following the locus $\Omega = \Omega_{\rm sta}$. This can be seen, for example, in Track 5, in the top left-hand panel of Fig.~\ref{darwindiagFG}. The track approaches the dotted line and follows it for a while. However, this is not a stable pseudo-equilibrium state, because there $\beta < 0$, but it can be maintained until the tidal torque becomes equal to the wind torque, $\dot{\Omega}$ reverses its sign, and $\beta$ becomes singular as given by Eq.~(\ref{eqldot}). When $\dot{\Omega} \geq 0$, the tidal torque dominates the wind torque, and necessarily $\beta > 1$. This corresponds in Fig.~\ref{darwindiagFG} to the domain below or to the right of the dotted curve. The pseudo-equilibrium cannot be reached because it would mean $\Omega > n$. 

There are here two possible courses of evolution. On one hand, if the tidal torque rises faster than the increase in magnitude of the wind torque, i.e., $\beta \rightarrow 1^{+}$, the planet will start an almost horizontal evolution in the diagram falling towards the star. This can be seen for all the tracks of the bottom left-hand panel of Fig~\ref{darwindiagFG} below the dotted line. On the other hand, if the increase in the tidal torque and the increase in the magnitude of the wind torque are comparable so that $\dot{\Omega} \simeq 0$, the evolution will proceed at almost constant rotation frequency as long as the wind torque stays comparable to the tidal torque. For example, this is what happens for Track 7 in the bottom right-hand panel of Fig.~\ref{darwindiagFG}. The track starts in the region where $\dot{\Omega} < 0$, below the dotted curve. The tidal torque is at first rising faster than the wind torque and the beginning of the track is almost horizontal. But when the wind torque and tidal torque become comparable, the track follows the locus $\Omega = \Omega_{\rm sta}$. Eventually, when the tidal torque  overcomes the wind torque, the evolution resumes an almost horizontal trajectory until the engulfment of the planet.

In conclusion, we find that in all the cases the continuous loss of angular momentum due to the stellar wind braking prevents our system from maintaining the pseudo-equilibrium state because,  even if it is reached at some stage of its evolution, it will  eventually become unstable.

 \subsection{Characteristic timescales of evolution}\label{timedep}
 Our model is too simplistic to  accurately describe the angular momentum evolution of actual stars and planets. But we can consider that there are different stages of the evolution characterized by different  relative importance of the  tidal and the wind torques. Indeed, there are two different processes that impact the evolution of the system: the loss of angular momentum by the stellar wind and the transfer of angular momentum from the orbit to the spin of the star by tides. Their respective characteristic timescales can be estimated as
\begin{equation}\label{timew}
\subrm{t}{w}  =\left|\frac{\subrm{L}{\star}}{\subrm{\tau}{w}}\right|,
\end{equation}
where $\subrm{L}{\star}$ is the stellar spin angular momentum and $\subrm{\tau}{w}$ is the wind torque; and
\begin{equation}\label{timet}
\subrm{t}{t}  =\left|\frac{h}{\subrm{\tau}{t}}\right| ,
\end{equation}
where $h$ is the orbital angular momentum and $\subrm{\tau}{t}$ is the tidal torque. This allows us to set different characteristic timescales of evolution for different zones of the Darwin diagram. Timescale estimates are a very rough way of describing the evolution of the angular momentum exchanges, as already stressed by \citet{Barker2009}, and we consider here only approximate cases for simplicity. As discussed in the previous section, there are three typical regimes that could be encountered at some stages during the tidal evolution of typical hot Jupiters. Those three regimes correspond to the wind torque that either dominates, is comparable to, or is dominated by, the tidal torque. We treat them here in this order, but they do not necessarily all happen for all possible exoplanetary systems. This depends on the initial distribution of angular momentum in the system. Furthermore, the  stationary rotation rate $\subrm{\Omega}{Sta}$ and all the thresholds introduced to separate different regimes  are sensitive to our model assumptions and parameter choices. Still, those timescales can be used to infer trends that can be tested against observations.

We now consider the phases of evolution that are mainly dominated by the wind AML. When the wind torque is much greater in amplitude than the tidal torque,  we can consider that the stellar spin sets the pace of evolution, as long as the ratio of tidal torque to wind torque is not greater than the ratio of orbital to rotational angular momentum (i.e $\subrm{t}{w}~<<~\subrm{t}{t}$ from Eqs.~\ref{timew} and \ref{timet}). This is the case for typical stellar rotation rates of young stars and planets not closer than the 2:1 mean motion resonance. For moderate rotators, the stellar spin is evolving with the characteristic timescale $\subrm{\tau}{w}~=~\subrm{\alpha}{mb}^{-1} \Omega^{-2}$. For stellar rotation periods of about 7, 10, and 30~days, this corresponds to about 0.5~Gyr, 1~Gyr, and 10~Gyr, respectively, for G-type stars (10 times longer for F-type stars). For faster rotators, the saturation of the wind must be accounted for and $\subrm{\tau}{w}~=~\subrm{\alpha}{mb}^{-1} \Omega^{-1} \subrm{\Omega}{sat}^{-1}$. If we take $\subrm{\Omega}{sat}~=~5.5\Omega_\odot$ \citep{Spada2011}, this yields values of $\subrm{\tau}{w}$ of about 50~Myr and 200~Myr for rotation periods of one and four days for G-type stars (again 10 times longer for F-type stars). Extremely close-in planets around very fast rotators, if formed, would have $\subrm{t}{t}~<<~\subrm{t}{w}$, so a different evolutionary timescale would apply.

\begin{figure}
	\centering
  	\includegraphics[width=\hsize]{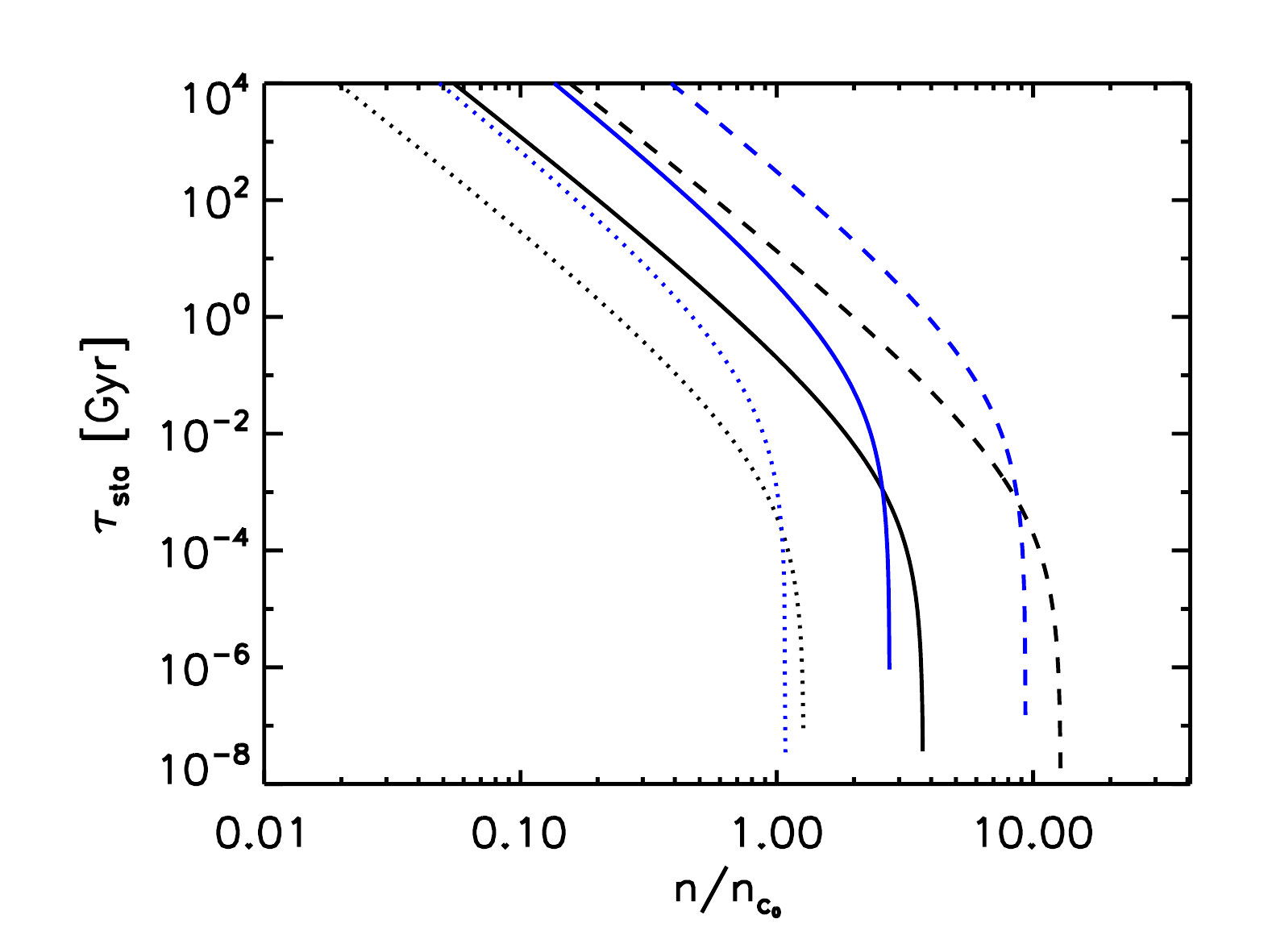}\\
  	\includegraphics[width=\hsize]{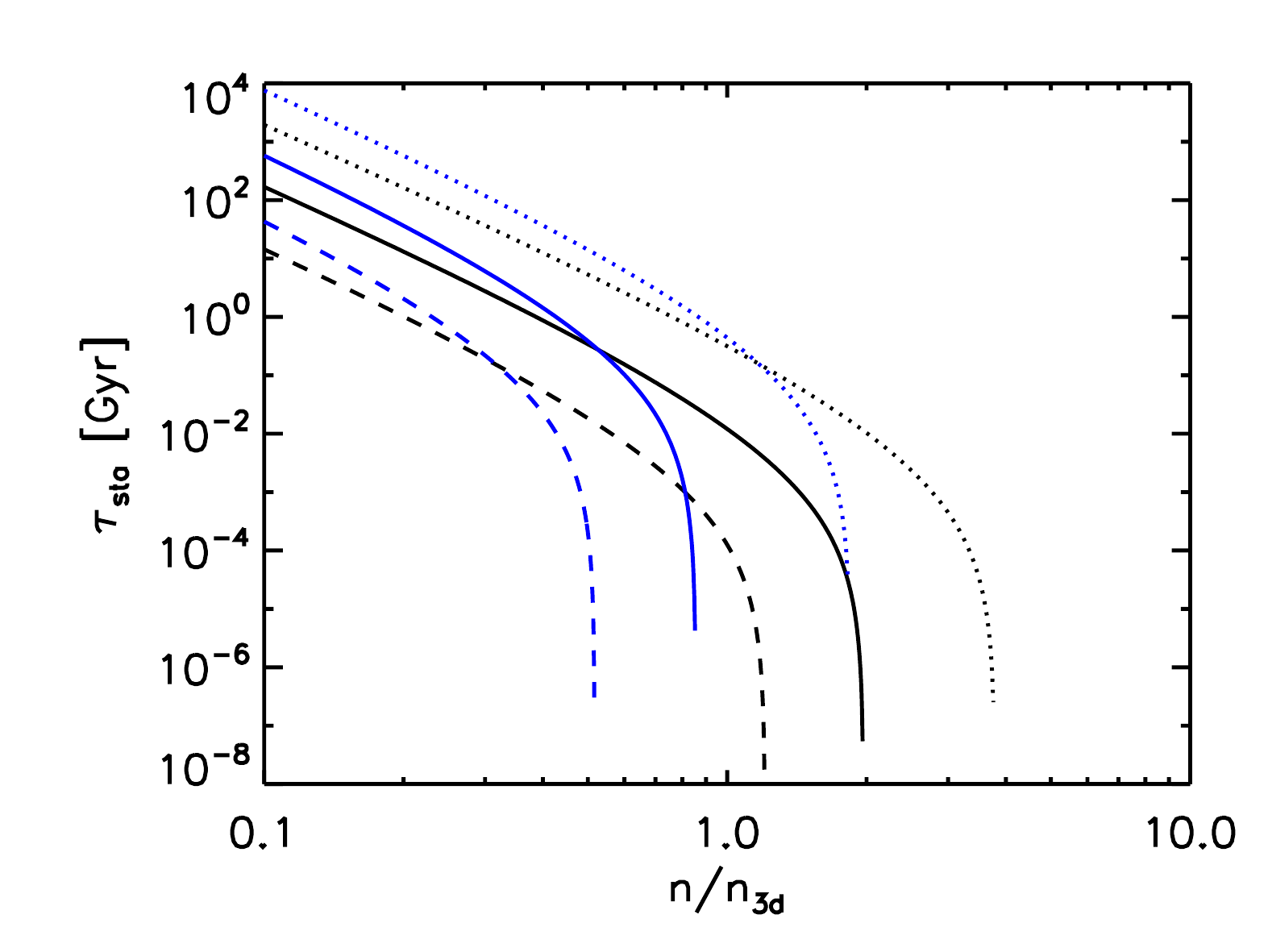}
      	\caption{Top: Estimates of the maximum possible duration of the stationary state $\subrm{\tau}{sta}$ as a function of the  orbital mean motion in units of the critical mean motion $\subrm{n}{c_0}$. The computations were done for a G-type star (black) or F-type star (blue) and planetary masses of  0.1 (dashed), 1 (solid), and 10 (dotted) Jupiter masses. We used $Q\rq{}=10^7$ for both G- and F-type stars, while the magnetic braking  coefficient $\subrm{\alpha}{mb}$ is reduced by a factor of ten  for F-type stars (see text). Bottom: The same, but the orbital mean motion is in units of $\subrm{n}{3d} = 2\pi/(3 {\rm days}$), hence independent of planetary and stellar mass.}
        \label{tausta}
\end{figure}

When the wind torque is comparable to the tidal torque, the system can enter a stationary state where tidal evolution proceeds at almost constant stellar spin frequency, which allows slowing down the migration of the planet. A necessary condition for the establishment of the stationary state is that the tidal torque be opposite in sign and comparable in magnitude to the wind torque, and this can be maintained as long as there is enough orbital angular momentum compared to the stellar rotational angular momentum to maintain the torque balance. Therefore, a rough estimation of the minimum possible duration of the stationary state $\subrm{\tau}{sta}$ is given by  
\begin{equation}
\subrm{\tau}{sta}=\frac{\Delta L}{\dot{L}}
\end{equation}
where $\Delta L= \subrm{L}{sta}-\subrm{L}{sta_c}$ is the excess of total angular momentum over the minimum value allowing the existence of torque balance, and $\dot{L}$ is the angular momentum loss rate corresponding to the value of $\Omega$ at the beginning of the stationary phase assumed to remain constant. Using Eqs.~\ref{eqstat} and~\ref{eqLtotLc2}, we can estimate $\subrm{L}{sta_c}$ and compute the corresponding duration of the stationary state as a function of the initial mean motion when a system enters into the stationary state, given in Fig.~\ref{tausta} for different stellar and planetary masses.

Stars losing less angular momentum through their wind (F-type stars) can generally maintain the stationary state longer than stars with a more efficient wind. For a given orbital distance, more massive planets can remain in the stationary state longer than less massive planets. However, the existence of the stationary state is limited to a maximum value of $n/\subrm{n}{c_0}$, which decreases for increasing mass. For Jupiter-sized planets , the stationary state cannot be maintained when $n\gtrsim 3.7 \subrm{n}{c_0}$, while massive planets cannot  maintain their stationary state when $n\gtrsim 1.3 \subrm{n}{c_0}$. In some cases, the stationary state can be maintained for a timescale longer than the main-sequence lifetime of the star. For example, this would be the case of a 10~M$_{\rm J}$ planet entering the stationary state with $n \lesssim 0.15-0.3 \subrm{n}{c_0}$ depending on the mass of the host star, which represents orbital periods greater than five to six days. For a Jupiter-sized planet, this would be the case if it starts with  $n \lesssim 0.4-0.9 \subrm{n}{c_0}$, which is about a 12-15~day orbital period. Finally, lighter planets can remain in the stationary state for tens of Gyrs as long as they enter it when $n \lesssim 1-3 \subrm{n}{c_0}$ depending on the mass of the host, which corresponds to orbital period greater than 20 days.

The final stages of the evolution see the planet spiralling into the star. This part of the evolution happens at almost constant angular momentum. We can calculate the tidal in-spiral time using the usual formula \citep[see e.g.][]{Barker2009}:
\begin{align}\label{eqtaua}
\tau_a &\equiv -\frac{2}{13}\frac{a}{\dot{a}}\\
  &\simeq 7 {\rm Gyr} \left(\frac{Q^\prime}{10^6}\right)\left( \frac{\subrm{M}{\star}}{\subrm{M}{\odot}}\right)^{1/2} \left( \frac{\subrm{M}{Jup}}{\subrm{M}{p}}\right)^{17/4} \left( \frac{\subrm{R}{\odot}}{\subrm{R}{\star}}\right)^{5}\nonumber\\
& \qquad \qquad \qquad\qquad \qquad  \left( \frac{\subrm{C}{\star}}{\subrm{C}{\odot}}\right)^{13/4} \left( \frac{n}{\subrm{n}{c_0}}\right)^{-13/3}  \left(1- \frac{\Omega}{n}\right)^{-1} .
\end{align}
To enter the final phase where the planet spiral inwards, we must have $\dot{\Omega} > 0$ and $L<\subrm{L}{sta_c}$. For a Jupiter-size planet in orbit around a solar-type star this means $n>3.7 \subrm{n}{c_0}$ which corresponds to $0.1 \leq \frac{\Omega}{n} \leq 0.5$ as long as $a<\subrm{a}{R}$, so $1<\left(1- \frac{\Omega}{n}\right)^{-1} < 2$. The in-spiral time is dominated by $\left( n/\subrm{n}{c_0}\right)^{-13/3}$; this implies $\tau_a \leq 100$~Myr when $n>3.7 \subrm{n}{c_0}$ for $Q\rq{}=10^7$. 

More massive planets enter the final phase of evolution for lower values of $n/\subrm{n}{c_0}$ and greater values of $n/\Omega$, but the in-spiral time is dominated by the factor containing the planet mass. Typically, for planets of more than 5~M$_{\rm J}$, this results in a in-spiral time $\tau_a $ of the order of  Myrs.  There is thus a very low probability to observe massive planets in this phase of evolution.  On the other hand, low-mass planets must have $n\gtrsim 10 \subrm{n}{c_0}$ to enter the tidally dominated phase of evolution, which can compensate for the effect of the mass term. However, this corresponds to values of  $\frac{\Omega}{n} \geq 0.9$, consequently their in-spiral time can be longer than the main-sequence lifetime of their host star. 

If tidal dissipation is stronger than what is considered here, since $A$ scales with $Q^\prime$, this will decrease the value of $A$ that will in turn decrease the ratio $n/\subrm{\Omega}{sta}$. In other words, for a given orbital distance, a stronger tidal dissipation would produce a stronger tidal torque, and the balance with the wind torque would be reached for a higher  rotation rate. However, $\tau_a$ is directly proportional to $Q\rq{}$, whereas the factor $\left(1- \frac{\Omega}{n}\right)$ does not increase as fast. Therefore, for given stellar and planetary masses, a stronger tidal dissipation leads to a smaller $\tau_a$.
\section{Discussion}\label{discussion}
The evolution of the semi-major axis of exoplanetary orbits results from the interplay between tides and magnetic braking of the host stars. As shown in Sect.~\ref{stabHJ}, the observed position of a given system in the Darwin diagram can help us determine its future evolution, but the duration of this evolution depends on the position of the system relative to the stationary state. As recalled in Sects.~\ref{tidtheo} and ~\ref{magbrak}, the relationship between the stellar and/or planetary  parameters and the efficiency of tidal dissipation and magnetic braking  is not well known, and the determination of the stationary locus for a given system cannot be accurate. However,  we expect remarkable differences between F- and G-type stars, which can be exhibited by the observation of general trends at the level of populations. Here, we discuss how the distribution of the known exoplanetary systems in a Darwin diagram can first, give us information about magnetic braking in F- and G-type stars and second, information about the initial orbital and rotational periods at the beginning of the tidal evolution.

\subsection{Trends in the Darwin diagram}
\begin{figure*}
	\centering
   	\includegraphics[width=0.8\hsize]{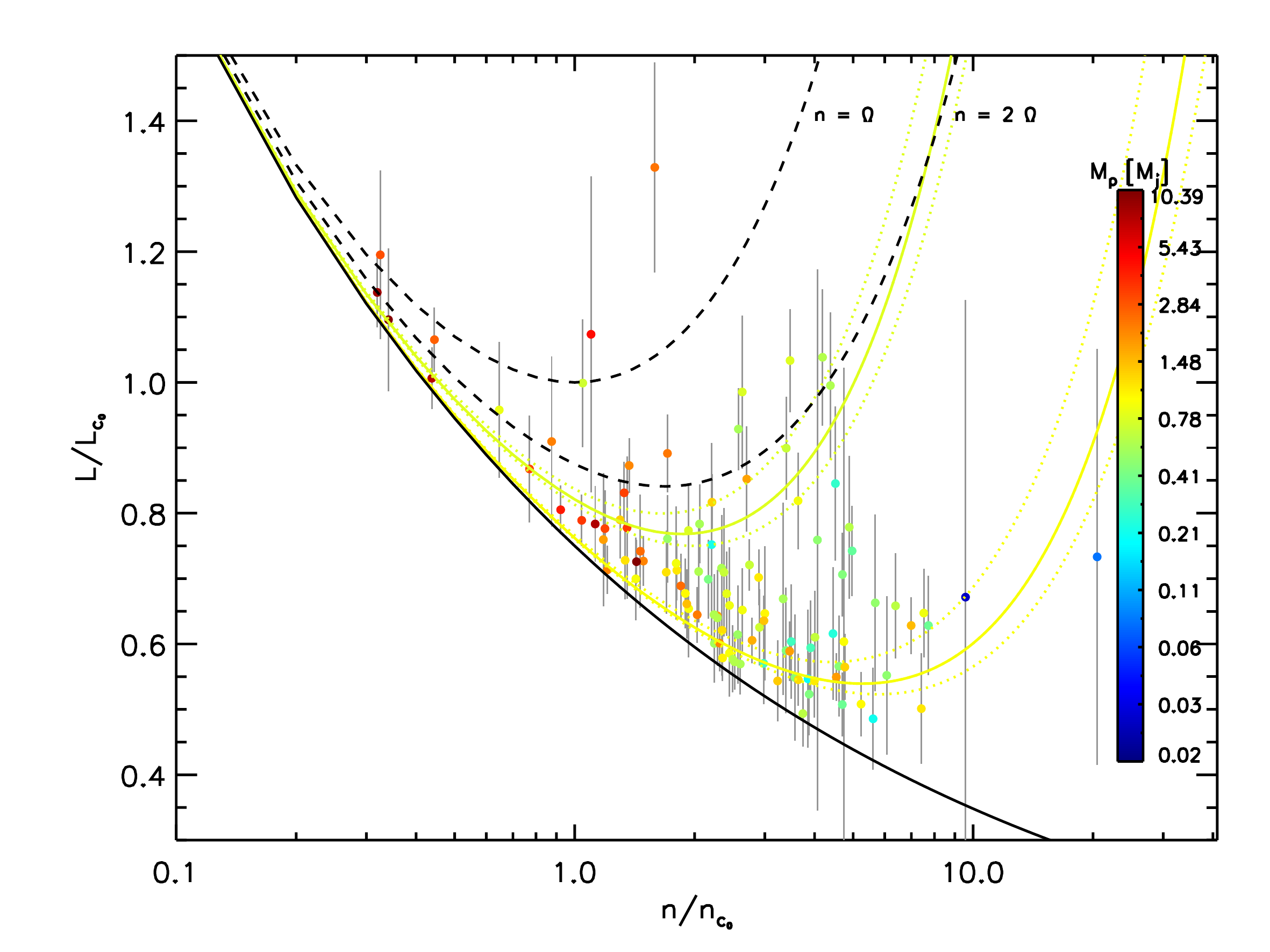}
      	\caption{Same as Fig. \ref{darwindiag}, but the colour of the symbols  indicates the mass of the planet. The dashed black lines indicate the loci where $n=\Omega$ and $n=2\Omega$ as labelled. The solid  black line is, for a given orbit, the contribution of the orbital angular momentum to the total momentum. The solid coloured lines are the best fits for the stationary locus for planets with $\subrm{M}{p} \leq 1$~M$_{\rm J}$ orbiting either stars with $\subrm{T}{eff} \geq 6250 $ or  $\subrm{T}{eff} < 6000 $~K. Their colours correspond to the average planetary mass of the respective subsamples. The 90\% confidence interval is given by the dotted lines of the respective colours.}
 	\label{darwindiagMass}
\end{figure*}
\begin{figure*}
	\centering
	\includegraphics[width=0.8\hsize]{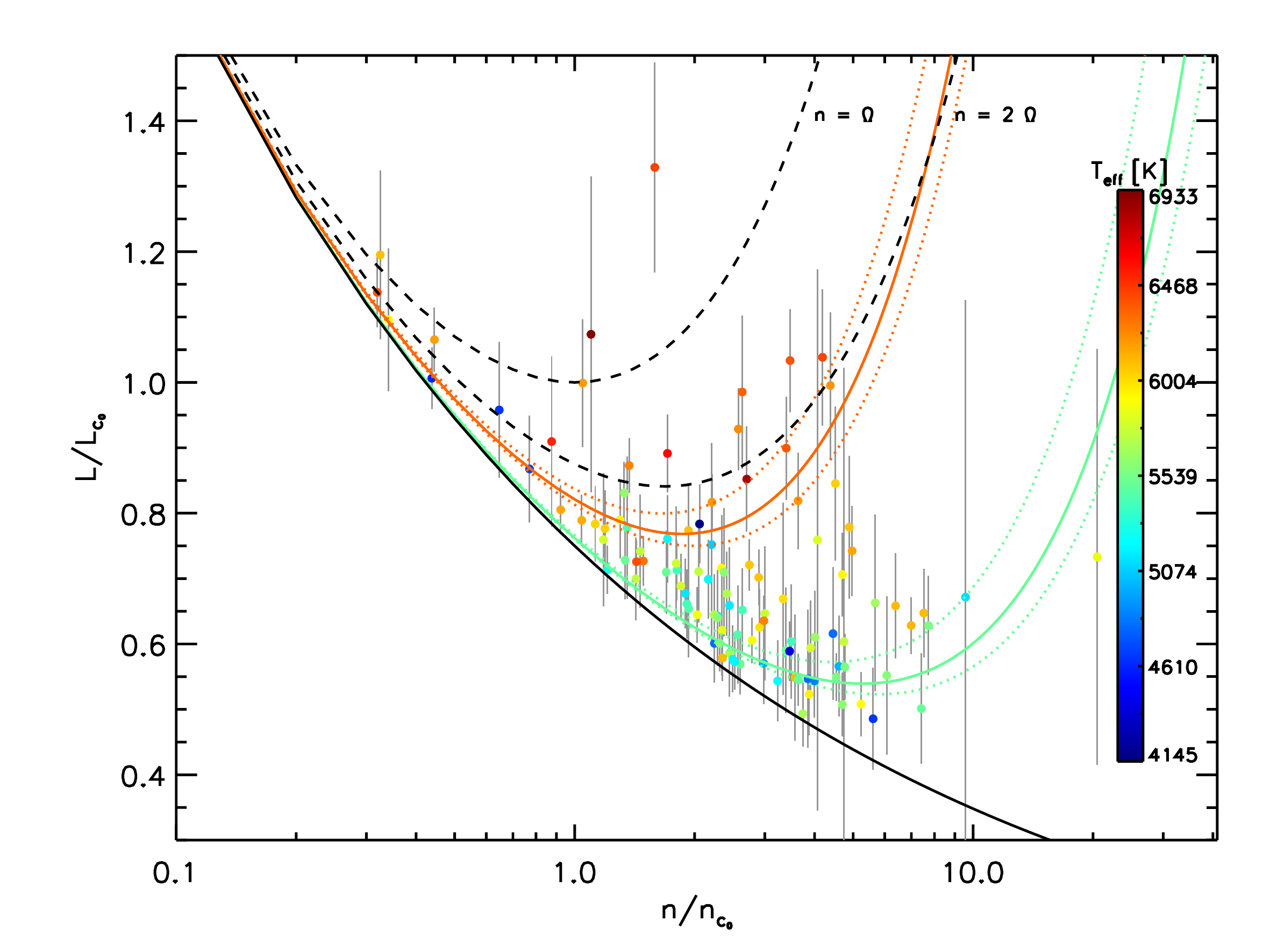} 
  	\caption{Same as Fig. \ref{darwindiagMass}, but the colour of the symbols indicates the effective temperature of the star. The colours of the solid lines give the mean effective temperature of the stars of both subsamples. The 90\% confidence interval is given by the dotted lines of the respective colours.}
	\label{darwindiagTeff}
\end{figure*}
    We again plot the same Darwin diagram as in Fig.~\ref{darwindiag}, but indicate the mass of the planet (Fig.~\ref{darwindiagMass}) and the effective temperature of the host star (Fig.~\ref{darwindiagTeff}) by the colour of the points.  By comparing both diagrams, it is clear that the position in the Darwin diagram depends not only on the mass of the planet, but also on the effective temperature of the star. We notice two important trends: firstly, in Fig.~\ref{darwindiagMass} we see that systems with more massive planets tend to have lower values of $n/\subrm{n}{c_0}$; secondly, in Fig.~\ref{darwindiagTeff} we see that, for a given orbital $n/\subrm{n}{c_0}$, systems with higher total angular momentum correspond to higher temperature hosts, and systems with excess rotational angular momentum do not appear to be uniformly distributed.  This is especially visible in the range $2 \leq n/\subrm{n}{c_0} \leq 5$, where systems with high host star temperature ($\subrm{T}{eff} \gtrsim 6250$~K) have a rotational angular momentum that contributes up to half of the total angular momentum.  But for $n/\subrm{n}{c_0} \gtrsim 5$, the rotational angular momentum excess is moderate, contributing to  about a third of the total angular momentum. We consider those two main trends in detail.
   
   \subsubsection{Planetary mass as a function of $n/\subrm{n}{c_0}$}
Since $\subrm{n}{c_0}$ is increasing with the planetary mass (Eq.~\ref{nc0}), the general trend of decreasing mass with increasing $n/\subrm{n}{c_0}$ mainly reflects that the planets in our sample have similar orbital periods (median value at 3.25 days and 95\% percentile at 5 days, the so-called \lq\lq{}pile-up\rq\rq{} at periods of $\sim3$~days; see e.g. \citealt{Gaudi2005}). For most systems, the orbital angular momentum accounts for at least three quarters of the total angular momentum  $L/\subrm{L}{c_0}$. This results in the general trend of higher total angular momentum with higher planetary mass. The pile-up of Jupiter mass planets at three days period has been widely discussed \citep{Cumming2008}, and its origin is still an open question. Here, we only notice that the stationary state cannot be maintained at greater $n/\subrm{n}{c_0}$ for higher planetary masses. According to Fig.~\ref{tausta}, for $ n/\subrm{n}{c_0}\gtrsim 1.5$, planets with $\subrm{M}{p} \gtrsim 3$~M$_{\rm J}$ cannot be in the stationary state regardless of the efficiency of magnetic braking of their host stars. They would be in the final stages of evolution where both the tides and the wind act to precipitate the planet into the star. For this mass range, this phase is so short that there is a very low probability of observing planets at those orbital distances. As $\subrm{n}{c_0}$ increases with increasing mass, this also means that the actual value of $n$ corresponding to the end of the stationary state depends weakly on the mass of the planet. As can be seen in Fig.~\ref{tausta}, for a Skumanich-type braking law and a tidal dissipation efficiency $Q\rq{}=10^7$, the end of the stationary state corresponds to $n / \subrm{n}{c_0} \approx 10, 3, $ and~1~for a 0.1, 1, and 10 Jupiter mass planet, respectively. This roughly corresponds to orbital periods of about 3-6, 1.5-4, and 1-2 days, respectively.
 
Finally, we must stress that the mean critical orbital period $\bar{\subrm{P}{c_0}}$ of our sample is $\sim5.85$ days. For single-site ground-based surveys, the probability of detecting a transit is close to one only for $P\leq2$~days. Multi-site surveys, such as HATNet and HAT-South, can approach five or six day completeness periods for very deep transits. This means that planets with $n<\subrm{n}{c_0}$ are not well detected and characterized with the current instruments \citep{Rauer13, Walker2013}. The Kepler mission might help to probe the orbital distribution to greater values, but the confirmation of the nature of the candidates from the ground remains a bottleneck. Examining the distribution of the orbital periods of Kepler planetary candidates nonetheless reveals that Jupiter-size candidates are indeed common at periods over ten days but only in multiple systems. On the contrary, the distribution of single Jupiter-size companion show a clear cut-off at around ten days. Unfortunately, it is not yet possible to quantify the distribution of semi-major axis as a function of the planetary mass for the single-planet Kepler candidates, because only the detected period can be considered with some confidence.

    \subsubsection{Systems with excess rotational angular momentum}  
\begin{figure}
	\centering
   	\includegraphics[width=0.8\hsize]{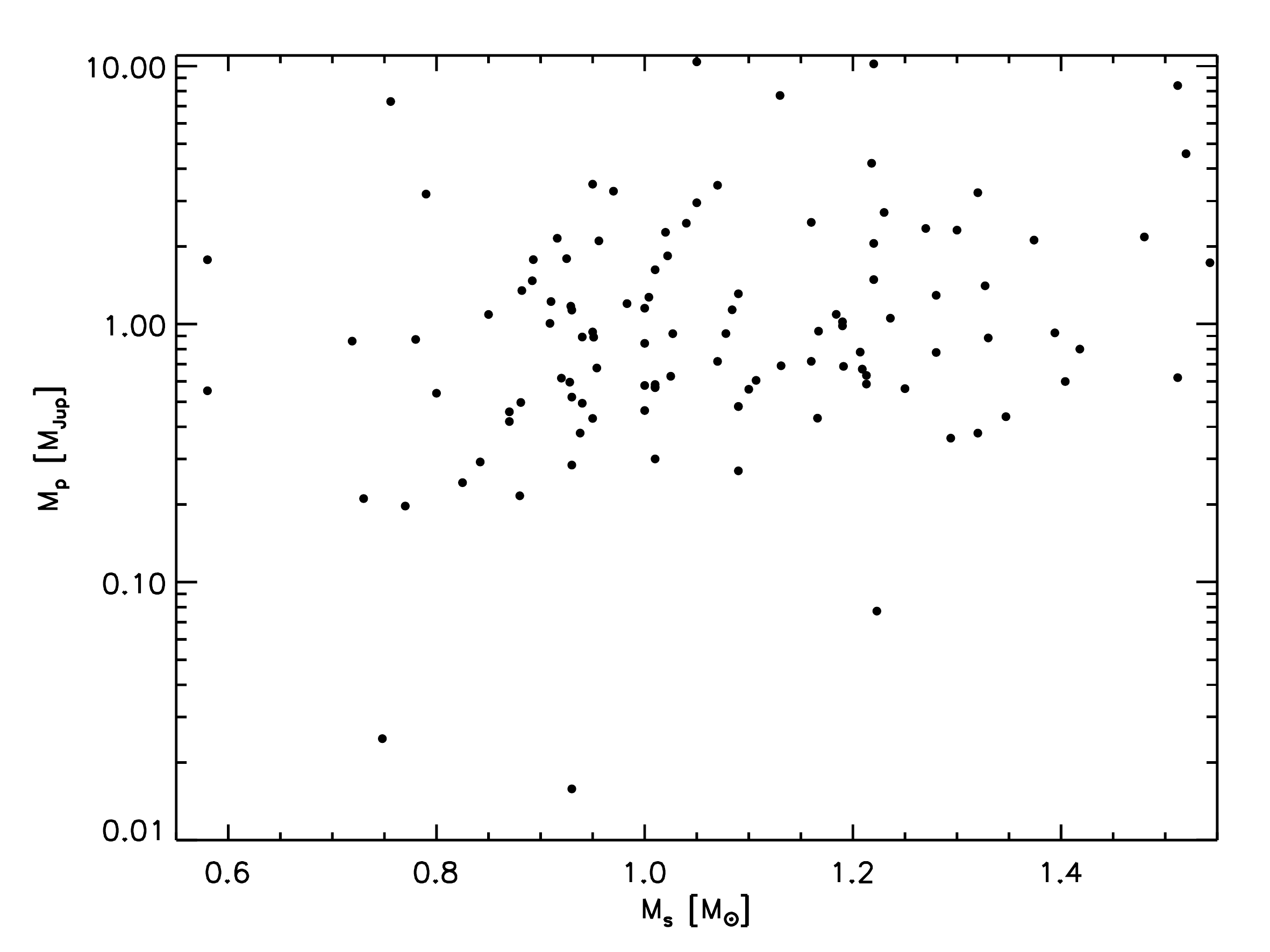} 
      	\caption{Planetary mass, in Jupiter mass, against the mass of the host, in Sun mass for the systems considered in this analysis. }
        \label{mplanmstar}
\end{figure}

Hotter stars generally rotate faster than cooler stars, given that they lose their angular momentum less efficiently. A system with a hotter star  will thus have proportionally more rotational angular momentum and reach a higher value of $L/\subrm{L}{c_0}$. This explains easily why systems with higher total angular momentum generally have higher temperature hosts for any given $n/\subrm{n}{c_0}$. While most systems have about one-fourth of their total angular momentum in the form of rotational angular momentum, systems with high temperature hosts ($\subrm{T}{eff} \gtrsim 6250$~K), in the range $2 \leq n/\subrm{n}{c_0} \leq 5$,  have a rotational angular momentum that contributes up to half of the total angular momentum. They show an excess of rotational angular momentum. But for $n/\subrm{n}{c_0} \gtrsim 5$, the rotational angular momentum excess is moderate, contributing to about one third of the total angular momentum.  In Fig.~\ref{darwindiagMass}, we also see that the systems that have a large excess of rotational angular momentum  (in the range $2 \leq n/\subrm{n}{c_0} \leq 5$) are homogeneous in planetary mass, which ranges between $\sim$0.6 and 0.8 M$_{\rm J}$, while the lower $L/\subrm{L}{c_0}$ cluster of systems in this range of $n/\subrm{n}{c_0}$ spans a broader mass range between $\sim$0.2 and 1.5~M$_{\rm J}$. This is a property that is partly shared with the whole sample. On one hand, the distribution of planet mass as a function of the host mass is homogeneous for planets that are more massive than 1~M$_{\rm J}$. On the other hand, the minimum mass detected around F-type stars  seems to be higher than for G-type stars (Fig.~\ref{mplanmstar}). 

A selection effect plays a role here, since planets are more difficult to detect and characterize around fast-rotating stars. For $n/\subrm{n}{c_0} \gtrsim 5$ and for planets more massive than about 1.0~$\subrm{M}{J}$, the rotation period $\subrm{P}{L/2}$ required to have at least half of the total angular momentum in the form of rotational angular momentum is less than about four to six days. This corresponds to $v \sin i > 8-10$~km/s for late-type stars, a value that could hinder the mass determination of planets in this mass range \citep[see e.g. ][]{Santerne2012} and affect the completeness of the surveys. For a lower mass planet, $\subrm{P}{L/2}$ increases and sets weaker constraints on radial velocities. Moreover, for a given mass, since the orbital angular momentum decreases for increasing $n/\subrm{n}{c_0} \gtrsim 5$, $\subrm{P}{L/2}$ increases with $n$. Thus, for a given mass, if we are able to detect a planet around a fast-rotating star having as much rotational angular momentum than the orbital angular momentum of the planet, we could also detect a planet of the same mass on a smaller orbit, with the same orbital-to-rotational angular momentum ratio, corresponding to a star not rotating as fast. Thus, for $n/\subrm{n}{c_0} > 5$ there seems to be a lack of systems with planetary mass lower than $\sim1$~$\subrm{M}{J}$ and excess rotational angular momentum that cannot be explained by observational biases alone.

This cannot be explained by tides alone either.  Indeed, in this part of the diagram $(n/\subrm{n}{c_0} \gtrsim 2$ and $L/\subrm{L}{c_0} \lesssim 1$), if we neglect magnetic braking, $\dot{\Omega} >0$ necessarily  and the planets would follow horizontal paths, leading to their tidal disruption within one or two characteristic timescales $\tau_a$. And yet, everything else being the same, the tidal in-spiral time $\tau_a$ is only about 1.5 times greater for an F- than for a G-type star (see Eq.~\ref{eqtaua}). Considering that F-type stars rotate faster, $\tau_a$ becomes about three times greater at the same $n/\subrm{n}{c_0}$  for an F-  than for a G-type star, because of the effect of the synchronization ratio  $\Omega/n$ on $\tau_a$. The in-spiral time is thus dominated by the planetary mass. Focusing on systems with $\subrm{M}{p} \leq 0.8 \subrm{M}{J}$ that should be free of observational biases,  and provided that the same population of planets is initially formed around F- and G-type stars, we should observe as many systems with rotational angular momentum excess before and after $n/\subrm{n}{c_0} > 5$. Actually, since F-type stars have a shorter main-sequence lifetime than G-type stars, tidal destruction would be more efficient for the latter, and we should not observe as many G stars at the same $n/\subrm{n}{c_0}$ when comparing F- and G-type hosts. This trend would be reinforced if we  consider that tidal dissipation efficiency could be an order of magnitude greater for G-type stars. 

In contrast, if we consider that the planet can enter the stationary state and delay its tidal evolution, the distribution of systems with excess angular momentum can be explained. For a Skumanich-type braking law and a tidal dissipation efficiency $Q\rq{}=10^7$, the stationary state can be entered for orbital periods of more than 1.5~days for a one Jupiter-mass planet orbiting a G-type star (respectively, $P\gtrsim3.5$~days  when orbiting an F-type star). The same reasoning for planets with $\subrm{M}{p}=10$~M$_{\rm J}$ shows that they can enter the stationary state if their initial period is more than 0.8~days for G-type stars (respectively $P\gtrsim1.6$~days  when orbiting an F-type star). Finally, less massive planets with $\subrm{M}{p}=0.1$~M$_{\rm J}$ have critical periods of 30-50 days, thus they can enter the stationary state for periods $P\gtrsim2.5$~days (respectively, $P\gtrsim5.8$~days  when orbiting an F-type star).

Most of the known planets have orbital periods longer than one day and could be in or near the stationary state. We computed the value of $\subrm{\alpha}{mb}$ and $Q\rq{}$ that would be required to observe those planets close to their stationary locus, i.e. with $\Omega \approx \subrm{\Omega}{sta}$. Since those two parameters are expected to depend on the spectral type of the host, we divided the sample in two subsamples, depending on the $\subrm{T}{eff}$ of the star (systems with $\subrm{T}{eff} \geq 6250$~K and systems with $\subrm{T}{eff}<6000$~K). The location of the stationary state also depends on the planetary mass, and we selected planets with $0.7 \leq \subrm{M}{p} \leq 1.1$~M$_{\rm J}$ to ensure statistically relevant sample sizes. Using Eqs.~\ref{eqstat} and \ref{eqLtotLc2},  we computed the value $A$ that best fits the stationary locus for the two subsamples.  From $A$ and using Eq.~\ref{eq:a}, we get the value of the product $\subrm{\alpha}{mb}Q\rq{}$ for each system.  We computed the mean to  account for a possible dispersion of the systems around the stationary state owing to their spread in age. Using a t-test, we find that the two means of the subsamples are different at the five percent level. Let $\subrm{\alpha}{mb} = \gamma 1.5 \times 10^{-14}$~yr and $Q\rq{}=\xi 10^7$. We cannot independently find the value of $\gamma$ and $\xi$, but only their product. We find a mean value of $\langle \gamma\xi \rangle = 2.5\pm 1.5$ for the subsample of cool host stars and $\langle \gamma\xi \rangle = 0.01 \pm   0.005$ for the hotter host stars. These values as well as their confidence intervals, are used to plot the locus of the stationary state represented in colours in Fig.~\ref{darwindiagTeff} and in Fig.~\ref{darwindiagMass}. The colours indicate the average planetary mass value in Fig.~\ref{darwindiagMass}, or the mean effective temperature of the star  in Fig.~\ref{darwindiagTeff} . If we assume that the tidal dissipation efficiency is the same for F- and G-type stars ($Q\rq{} =10^7$), we find the expected order of magnitude for the magnetic braking coefficient for G-type stars, but about a factor of ten smaller than  expected for F-type stars. However this approach is very crude, and it is possible that some of the systems have already gone past the stationary state which would increase the  estimated value of the product $\gamma \xi$. Nevertheless, this analysis suggests that the combined effect of magnetic braking and tidal evolution could be significantly different depending on the effective temperature of the host star. Lastly, let us note that supposing that many systems could be near the stationary state only places a lower bound on the value of $\gamma \xi$. Indeed, if tidal dissipation is much weaker than considered here, in particular for the F-type stars, the position in the Darwin diagram could result from the formation process and AML of the stars only.

\begin{figure}
	\centering
  	\includegraphics[width=\hsize]{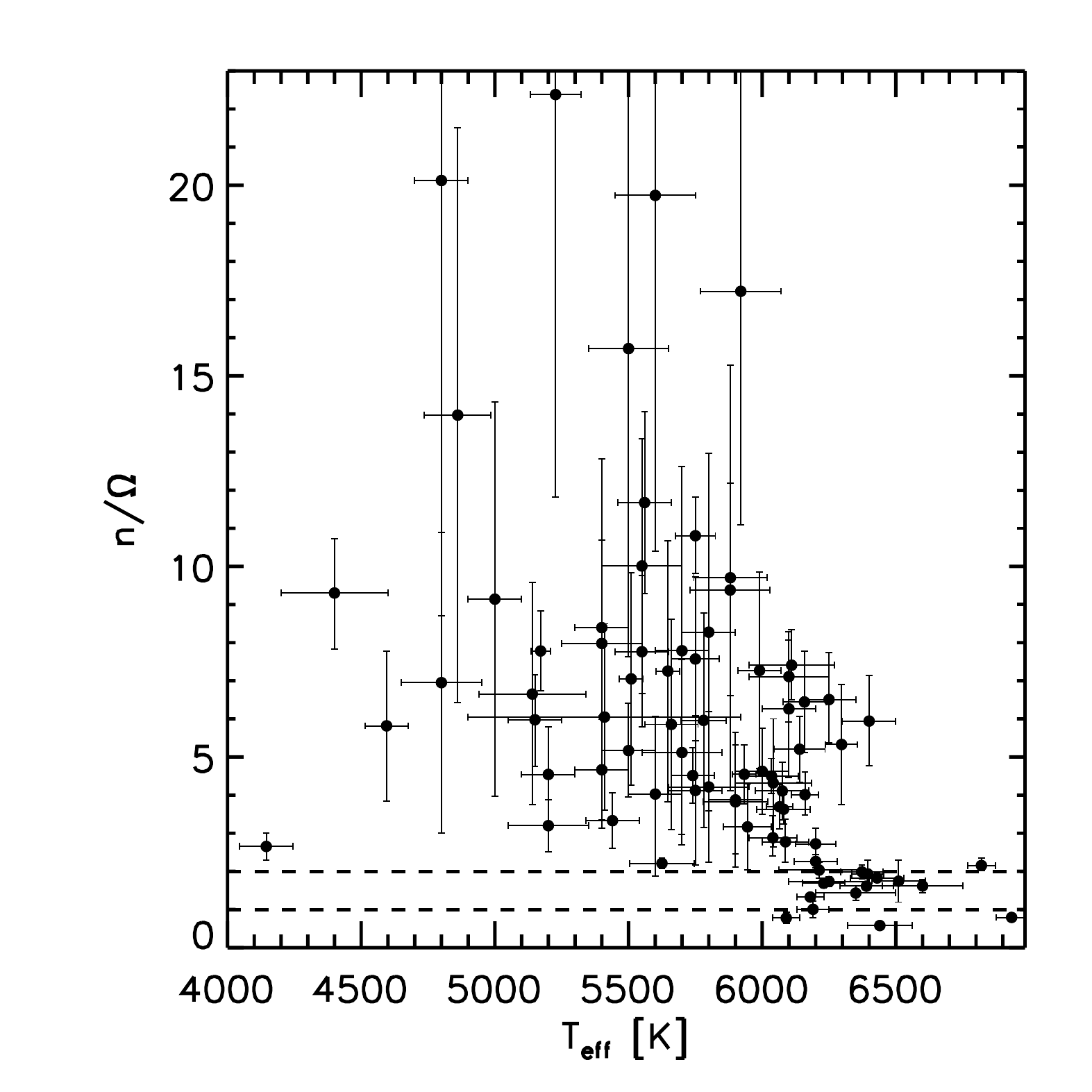}
	\caption{Ratio of the orbital mean motion to the stellar angular velocity vs. the effective temperature of the star in aligned and circular systems.}
	\label{synchteff}
\end{figure}

Moreover, we  notice that systems with excess rotational angular momentum have  $\subrm{T}{eff} \gtrsim 6200$~K and seem to cluster around the locus where $n=2\Omega$, while the others span a wider range of synchronization ratio $n/\Omega$. This can be clearly seen in  Fig.\ref{synchteff}, where the ratio $n/\Omega$ is displayed as a function of the effective temperature of the host star for the whole sample.
There is a clear difference between systems with host stars with $\subrm{T}{eff} \gtrsim 6000$~K, that all have $n/\Omega <8$, and those with cooler hosts that span a wide range of the ratio values, generally greater than 2, with a possible trend indicating higher value of $n/\Omega$ for lower effective temperature. Moreover, almost all systems with $\subrm{T}{eff} \gtrsim 6250$~K have $1<n/\Omega \leq 2$. With a smaller sample, \citet{Lanza2010} had already put forward this remarkable dependence. Now, with three times as much systems, we can definitely confirm this trend. If magnetic braking is indeed efficient in G-type stars, $\Omega$ decreases with time. The overall value of $n/\Omega$ becomes large, and for cooler stars, the spread in $n/\Omega$ reflects the spread in age among those systems. On the other hand, F-type stars loose very little angular momentum through their wind and are not efficiently braked. 
\begin{figure}[h]
	\centering
   	\includegraphics[width=0.49\hsize]{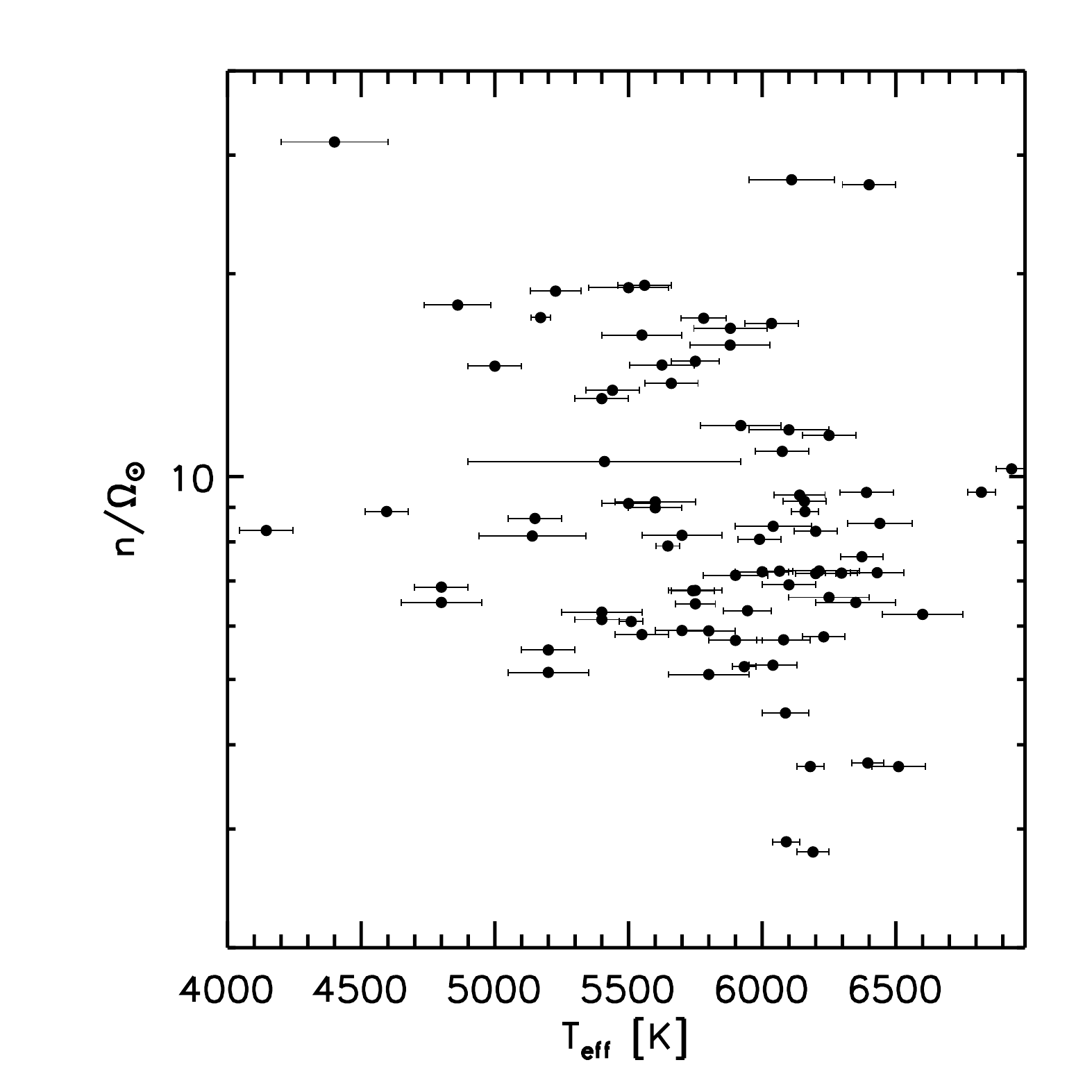}
   	\includegraphics[width=0.49\hsize]{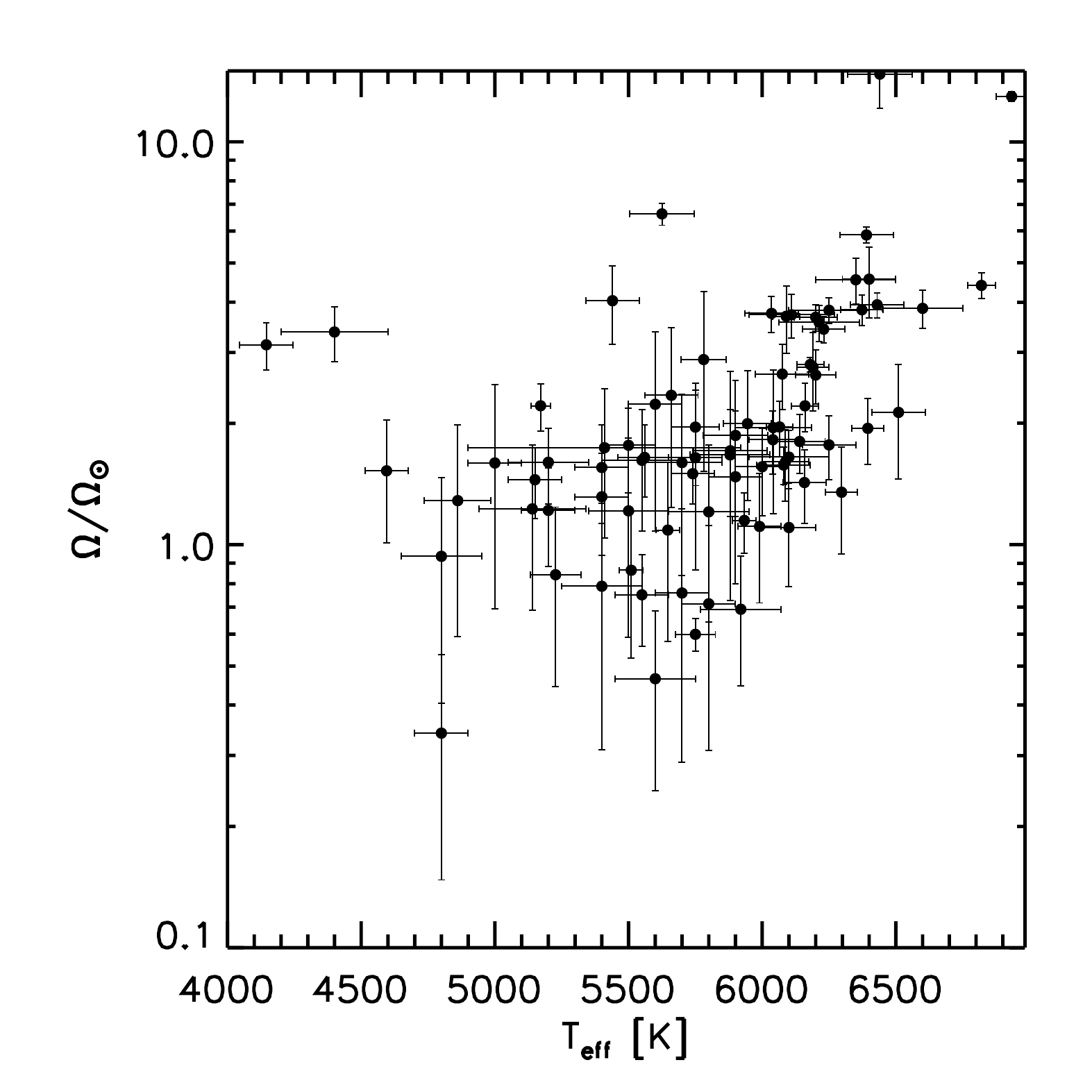}
      	\caption{Orbital mean motion (left) and stellar angular velocity (right) in units of the Sun\rq{}s angular velocity vs. the effective temperature of the star in aligned and circular systems.}
	\label{norbvsteff}
\end{figure}  
   We plot in Fig.~\ref{norbvsteff} (right) the observed values of $\Omega$ vs. the effective temperature for the stars of our sample. We observe that $\Omega$ is indeed homogeneous for F-type and larger than for G-type stars, in agreement with a behaviour that is dominated by magnetic braking.  At the same time, a plot of $n$ vs. the stellar effective temperature  (Fig.~\ref{norbvsteff}, left) does not show any remarkable trend. Thus, for F-type stars, the stellar rotation is fast enough for the wind torque to dominate the tidal torque, while at the same time magnetic breaking is weak enough as to not significantly change the stellar rotation rate. This means that for F-type stars, the value of $n/\Omega$ that we observe today could be similar to the one attained by the system when it began its tidal evolution.  
   
\subsection{Inferring past evolution}
In this way, the current position of a system in the Darwin diagram not only gives information about its future evolution, but also constrains its initial orbital and rotational periods. Depending on the main migration mechanism, the beginning of the tidal evolution, what we call the initial state, occurs at different ages after the formation of the system. In any case, the Darwin diagram assumes constant stellar and planetary mass and radius, so our approach is relevant only once the star has settled on the zero-age main sequence (ZAMS). 

Let us first assume that the main migration ended by the time the star reached the main sequence, as would be the case for a disk-driven migration. In this case, planets should halt their migration where the differential Lindblad torque reduces to zero \citep{Kuchner2002}, which corresponds to the location of the 2:1 mean motion resonance with the edge of the inner cavity of the disk. Smaller planets might also be trapped slightly outside of the disk inner cavity \citep{Benitez-Llambay2011}. Supposing that the strong magnetic field of the star can disrupt the inner disk region, we  expect that the inner disk edge is co-rotating with, and lies at a few stellar radii from, the central star \citep{Bouvier2007}. This means that we should have $n\approx \Omega$ for low mass planets, and $n=2\Omega$ for a more massive planet when the disk disappears, typically within the first 5-10~Myr of pre-main-sequence evolution \citep{Hernandez2010}. At this stage, most late-type stars have rotational periods between two and ten days \citep{Gallet2013}, so the planets should have orbital periods between one and ten days. However, those stars are in the pre-main sequence phase of their evolution and are still contracting. 

Supposing that the contraction is much faster than tidal interactions, most planets would have $n/\Omega < 2$ when the star reaches the main sequence. If $n/\Omega < 1$, the planet then has $\subrm{P}{rot}<\subrm{P}{orb}<\subrm{P}{sta}$, thus it is driven into the stationary state with very little change in its mean orbital motion.  This phase can last for about 500~Myr for G-type stars, or 5~Gyr for F-type stars. When $\subrm{P}{rot}$ becomes comparable to $\subrm{P}{sta}$, and assuming that $\subrm{P}{orb}$ has not changed significantly, we have $\subrm{P}{orb}<\subrm{P}{rot}<\subrm{P}{sta}$, and most planets around G-type stars will  enter the stationary state. It can be maintained at most for over 1~Gyr for Jupiter-mass planets, and up to a few tens of Gyrs for more massive planets, while low-mass planets remain in the stationary state at most for a few 100~Myrs. On the other hand, F-type stars can only retain the more massive planets in the stationary state, since only small-mass planets with orbital periods longer than about 6~days can equal the wind torque with their tidal torque. For the same orbital period, the stationary state can then be maintained longer for greater mass: at most a few Gyrs for Jupiter-mass planets, and up to a few tens of Gyrs for more massive planets. The positions of known exoplanetary systems in the Darwin diagram agree at least qualitatively with this kind of temporal evolution.
 
In migration scenarios involving the secular interaction with a distant third body  so far that it does not take part directly in the tidal interaction (e.g. Kozai-Lidov induced migration), the close binary tidal evolution starts some time after the arrival of the star on the ZAMS. The third body cyclically excites the eccentricity of the orbit of the inner planet that can reach values $e \sim 1$. During one of those high-eccentricity excursions, the planet can experience a close tidal encounter with the star, and its orbit can be circularized at almost constant orbital angular momentum. The closest approach before circularization corresponds to a periastron distance $\approx a_{\rm R}$, thus the final semi-major axis after circularization will be  $a\sim2\subrm{a}{R}$ \citep{Rasio1996}. This corresponds to  $n/\subrm{n}{c_0}$ of about 1.7, 2.9, and 16.5 for planets of mass 10, 1, and 0.1~M$_{\rm J}$ around G-type stars, and of 2.9, 5.2, and 29.2 for planets of mass 10, 1, and 0.1~M$_{\rm J}$ around F-type stars. This is in reasonably good agreement with the observed orbital mean motion of some systems, but it cannot explain all the features of the Darwin diagram, in particular the absence of low-mass planets around F-type stars (which holds also for eccentric systems). Moreover, in this case, for a Skumanich-type braking law and a tidal dissipation efficiency $Q\rq{}=10^7$, those planets could not maintain the stationary state, regardless of the spectral type of the star. 

If at the end of the migration $\subrm{P}{orb}<\subrm{P}{rot}<\subrm{P}{sta}$, the evolution in the Darwin diagram would first be dominated by the wind torque. Its duration depends on the stellar rotation frequency and on the time scale required for the excitation of the eccentricity of the inner planet and its subsequent tidal damping. For circular and aligned systems, we can assume that this phase is shorter than the main-sequence lifetime of the star. Then, when $\subrm{P}{orb}<\subrm{P}{sta}<\subrm{P}{rot}$, the stationary state is surpassed and the planet can only dive into the star. For a Jupiter-sized planet, this corresponds to stellar rotation periods longer than about 16 days, for G-type star with a Skumanich-type braking law and a tidal dissipation efficiency $Q\rq{}=10^7$, and to about 9~days for F-type stars. The planet would then take no more than a few hundreds Myrs to reach its Roche limit. For much more massive planets, this corresponds to stellar rotation periods longer than about 1.0~day for G and F-type stars, and the planet would then take less than 1~Myr to reach its Roche limit. For lower mass planets, this corresponds to orbital periods below about two days, but at those orbital periods they have stationary stellar rotation periods greater than 40-70~days depending on the stellar mass, so they can remain in the wind-dominated part of the momentum evolution for most of the main-sequence lifetime of the star. In conclusion, except for the less massive planets, those systems would have a very short lifetime in orbit around their host star once they have been circularized. However, different values of tidal dissipation efficiency or, more importantly, magnetic braking could significantly increase their lifetime by putting them in the stationary state. 

A more accurate knowledge of the age of the systems, together with more detailed simulations, are required to put some constraints on tidal dissipation efficiency, magnetic braking, and migration scenarios.
\section{Conclusion}

We have rigorously demonstrated for the first time the possible outcomes of tidal interaction between a planet and a star when the magnetic braking of the star is considered. We showed that a pseudo-stable equilibrium state can exist that is characterized by circularity of the orbit, alignment between the spin of the planet, the star, and the normal to the orbital plane, but not by co-rotation. The orbital mean motion of the planet at equilibrium is equal to the stellar angular velocity reduced by a factor that depends on the angular momentum loss rate through the stellar wind. We proposed a set of dimensionless variables that allow representing the planets relative to their stationary equilibrium state in a single diagram, thereby extending a graphical method first introduced by \citet{Darwin1879}. In the stationary state, the tidal torque can compensate for the wind torque, resulting in a remarkably different  evolution of the angular momentum in exoplanetary systems around F- or G-type stars.  

We provided estimates for the characteristic timescales associated with the main phases of this  evolution.  For a Skumanich-type braking law and tidal dissipation efficiency $Q\rq{}=10^7$, we showed that the evolution of the angular momentum when the star is on the main sequence is mainly driven by magnetic braking, especially in the first Gyrs of the evolution of the system. In F-type stars, the wind torque is strong enough to dominate the tidal torque for close-in planets, but at the same time weak enough not to induce remarkable changes in their semi-major axis, over periods of a few Gyrs. We found a statistically significant difference  between the distributions of angular momentum in systems with F and G-type stars and discussed the possibility that most of the transiting planets in circular and aligned orbits are close to their stationary state. The current position in the Darwin diagram gives information not only about the future evolution of a given system, but also on  its possible initial angular momentum distribution at the beginning of binary tidal interaction. More detailed studies could help to put constraints on tidal dissipation efficiency, magnetic braking, and migration scenarios.  

Our approach assumes a constant tidal dissipation efficiency and a Skumanich-type braking law. As put forward by several authors, these approximations may not always be accurate. Moreover, we overlooked the pre-main-sequence (PMS) phase of stellar evolution, which can be crucial for testing migration theories. As explained in Sect.~\ref{tidtheo}, the efficiency of tidal dissipation could be very sensitive to stellar spin, so that from the PMS to the end of the main sequence, exoplanetary systems with late-type stars could experience very different amounts of tidal dissipation. Dedicated simulations are required to include the PMS phase. While our model is too simplistic to accurately describe the  evolution of a particular system, it is helpful to point out general trends in the populations of close-in companions. For example, recent results for the rotation periods of Kepler candidate host stars showing a dearth of close-in companions around fast-rotating stars \citep{McQuillan2013} may be explained by the effect of magnetic braking on tidal evolution \citep{Teitler14}. Indeed, for a wide range of planetary and stellar masses, the stationary state on such close-in orbits cannot be maintained. Moreover, the stationary state for planets on close-in orbits is generally reached at high stellar angular velocity. This means that the vertical path phase in the Darwin diagram for close-in planets around fast rotators is generally fast. When the tidal torque overcomes the wind torque, they quickly spiral towards their star  and are tidally disrupted.  

Our results agree with previous studies showing that the final fate of most known transiting exoplanets is to merge with their stars \citep{Rasio1996a, Levrard2009a, Matsumura2010, Jackson2009, Guillochon2011}. However, most of those studies estimated $Q\rq{}_\star$ from the orbital decay timescales obtained by considering tidal interaction alone, i.e., by neglecting the effects of magnetic braking. In particular, the study by \citet{Jackson2009} uses the lower envelope of the distribution of the semi-major axis and stellar age estimates to infer the best fitting $Q\rq{}_\star$. This can lead to underestimating the tidal dissipation efficiency of the star, if the planets have entered the stationary state in the past. However, this might be a good estimate for Jupiter-sized planets, as far as they started their tidal evolution so close to their host star to be unable to enter the stationary state and are probably already in the tidally dominated part of the evolution. Nevertheless, inferences on the initial semi-major axis distribution of close-in planets can be significantly affected by the existence of the stationary state. Moreveor, the existence of the stationary state implies that stars hosting close-in companions may not be as strongly braked by their wind as their counterparts without companions. Therefore, the gyrochronology law for exoplanets host must be calibrated on a suitable sample and take this effect into account \citep[see ][]{Pont2009,Lanza2010,Poppenhaeger14}.

In conclusion, stellar magnetic braking has to be considered to properly describe the tidal evolution of close-in exoplanets, especially on long timescales. Reliable ages and rotation rates of the host stars are required to put relevant constraints on tidal dissipation efficiency and migration theories. Since the mass of the planet is crucial for tidal effects, the faint planetary candidates of the Kepler mission offer limited possibilities for this kind of study. However, the PLATO mission, recently accepted as a class M mission for the ESA Cosmic Vision plan \citep{Rauer13}, will provide accurate stellar and planetary parameters, including ages, by the asteroseismic study of the host stars of transiting planets over a wide range of orbital periods and planetary radius. Focussing on bright targets, PLATO will allow their radial velocity follow-up and planetary mass determination. It will thus be possible to assess the initial distribution of the semi-major axis of close-in planets and constrain migration theories.
\begin{acknowledgements}
The authors are grateful to the anonymous referee for valuable comments on the manuscript that resulted in significant improvements. This research has made use of the Exoplanet Orbit Database
and the Exoplanet Data Explorer at exoplanets.org. CD acknowledges support by CNES grant 426808. 
\end{acknowledgements}
\bibliographystyle{aa}
\bibliography{Darwin_stab.bib}

\onecolumn
\begin{longtable}{lcccccccccc}

\caption{\label{tab:param} Main stellar and planetary parameters of the systems considered in Sec.\ref{stabHJ}. The errors on the orbital period and semi-major axis are not reproduced to save space but are typically smaller than the $10^{-5}$ days and $10^{-3}-10^{-4}$ AU, respectively.}\\
\hline\hline
Name& $\subrm{P}{orb}$ & $a$ & $e$ & $\subrm{M}{p}$ & $\subrm{R}{p}$  & $\subrm{M}{\star}$& $\subrm{R}{\star}$& $\subrm{T}{eff}$& $v \sin i$  & $\lambda$ \\
& (days) &(au) &  & (M$_{\rm J}$) &(R$_{\rm J}$) & (M$_{\odot}$) & (R$_{\odot}$)& (K) & (km.s$^{-1}$) & ($^\circ$) \\
\hline
\endfirsthead
\caption{continued.}\\
\hline\hline
Name& $\subrm{P}{orb}$ & $a$ & $e$ & $\subrm{M}{p}$ & $\subrm{R}{p}$  & $\subrm{M}{\star}$& $\subrm{R}{\star}$& $\subrm{T}{eff}$& $v \sin i$  & $\lambda$ \\
& (days) &(au) &  & (M$_{\rm J}$) &(R$_{\rm J}$) & (M$_{\odot}$) & (R$_{\odot}$)& (K) & (km.s$^{-1}$) & ($^\circ$) \\
\hline
\endhead
\hline
\endfoot
   CoRoT-11& 2.9943&0.044&0 & 2.35$^{ 0.34}_{- 0.34}$ & 1.43$^{ 0.03}_{- 0.03}$ & 1.27$^{ 0.05}_{- 0.05}$ & 1.37$^{ 0.03}_{- 0.03}$ &6440$\pm$120&40.0$^{ 5.0}_{- 5.0}$ & 0.1$^{ 2.6}_{- 2.6}$\\
   CoRoT-12& 2.8280&0.040&0.070$^{0.063}_{- 0.04}$ & 0.92$^{ 0.07}_{- 0.07}$ & 1.44$^{ 0.13}_{- 0.13}$ & 1.08$^{ 0.08}_{- 0.07}$ & 1.12$^{ 0.10}_{- 0.09}$ &5675$\pm$ 80& 1.0$^{ 1.0}_{- 1.0}$ &-\\
   CoRoT-13& 4.0352&0.051&0 & 1.31$^{ 0.08}_{- 0.08}$ & 0.89$^{ 0.01}_{- 0.01}$ & 1.09$^{ 0.02}_{- 0.02}$ & 1.01$^{ 0.03}_{- 0.03}$ &5945$\pm$ 90& 4.0$^{ 1.0}_{- 1.0}$ &-\\
   CoRoT-14& 1.5121&0.027&0 & 7.69$^{ 0.45}_{- 0.45}$ & 1.09$^{ 0.07}_{- 0.07}$ & 1.13$^{ 0.09}_{- 0.09}$ & 1.21$^{ 0.08}_{- 0.08}$ &6035$\pm$100& 9.0$^{ 0.5}_{- 0.5}$ &-\\
   CoRoT-17& 3.7681&0.048&0 & 2.46$^{ 0.28}_{- 0.28}$ & 1.02$^{ 0.07}_{- 0.07}$ & 1.04$^{ 0.10}_{- 0.10}$ & 1.51$^{ 0.05}_{- 0.05}$ &5740$\pm$ 80& 4.5$^{ 0.5}_{- 0.5}$ &-\\
   CoRoT-18& 1.9001&0.030&0.040$^{0.040}_{- 0.04}$ & 3.49$^{ 0.38}_{- 0.38}$ & 1.31$^{ 0.18}_{- 0.18}$ & 0.95$^{ 0.15}_{- 0.15}$ & 1.00$^{ 0.13}_{- 0.13}$ &5440$\pm$100& 8.0$^{ 1.0}_{- 1.0}$ &10.0$^{20.0}_{-20.0}$\\
    CoRoT-2& 1.7430&0.028&0.014$^{0.008}_{- 0.01}$ & 3.28$^{ 0.17}_{- 0.17}$ & 1.47$^{ 0.04}_{- 0.04}$ & 0.97$^{ 0.06}_{- 0.06}$ & 0.90$^{ 0.02}_{- 0.02}$ &5625$\pm$120&11.8$^{ 0.5}_{- 0.5}$ & 7.2$^{ 4.5}_{- 4.5}$\\
   CoRoT-25& 4.8607&0.058&0 & 0.27$^{ 0.04}_{- 0.04}$ & 1.08$^{ 0.30}_{- 0.10}$ & 1.09$^{ 0.11}_{- 0.05}$ & 1.19$^{ 0.14}_{- 0.03}$ &6040$\pm$ 90& 4.3$^{ 0.5}_{- 0.5}$ &-\\
   CoRoT-26& 4.2047&0.052&0 & 0.48$^{ 0.07}_{- 0.07}$ & 1.26$^{ 0.13}_{- 0.07}$ & 1.09$^{ 0.06}_{- 0.06}$ & 1.79$^{ 0.18}_{- 0.09}$ &5590$\pm$100& 2.0$^{ 1.0}_{- 1.0}$ &-\\
   CoRoT-27& 3.5753&0.047&0.000$^{0.065}_{- 0.00}$ &10.39$^{ 0.77}_{- 0.77}$ & 1.01$^{ 0.04}_{- 0.04}$ & 1.05$^{ 0.11}_{- 0.11}$ & 1.08$^{ 0.18}_{- 0.06}$ &5900$\pm$120& 4.0$^{ 1.0}_{- 1.0}$ &-\\
    CoRoT-4& 9.2020&0.090&0.000$^{0.100}_{- 0.00}$ & 0.72$^{ 0.07}_{- 0.07}$ & 1.19$^{ 0.06}_{- 0.06}$ & 1.16$^{ 0.03}_{- 0.02}$ & 1.17$^{ 0.01}_{- 0.03}$ &6190$\pm$ 60& 6.4$^{ 1.0}_{- 1.0}$ &-\\
    CoRoT-5& 4.0379&0.050&0.090$^{0.090}_{- 0.04}$ & 0.46$^{ 0.04}_{- 0.04}$ & 1.39$^{ 0.05}_{- 0.05}$ & 1.00$^{ 0.02}_{- 0.02}$ & 1.19$^{ 0.04}_{- 0.04}$ &6100$\pm$ 65& 1.0$^{ 1.0}_{- 1.0}$ &-\\
    CoRoT-6& 8.8866&0.085&0.000$^{0.100}_{- 0.00}$ & 2.95$^{ 0.33}_{- 0.33}$ & 1.17$^{ 0.04}_{- 0.04}$ & 1.05$^{ 0.05}_{- 0.05}$ & 1.02$^{ 0.03}_{- 0.03}$ &6090$\pm$ 50& 7.5$^{ 1.0}_{- 1.0}$ &-\\
    CoRoT-8& 6.2123&0.063&0 & 0.22$^{ 0.03}_{- 0.03}$ & 0.57$^{ 0.02}_{- 0.02}$ & 0.88$^{ 0.04}_{- 0.04}$ & 0.77$^{ 0.02}_{- 0.02}$ &5080$\pm$ 80& 2.0$^{ 1.0}_{- 1.0}$ &-\\
   HAT-P-12& 3.2131&0.038&0 & 0.21$^{ 0.01}_{- 0.01}$ & 0.96$^{ 0.03}_{- 0.02}$ & 0.73$^{ 0.02}_{- 0.02}$ & 0.70$^{ 0.02}_{- 0.01}$ &4650$\pm$ 60& 0.5$^{ 0.4}_{- 0.4}$ &-\\
   HAT-P-16& 2.7760&0.041&0.036$^{0.004}_{- 0.00}$ & 4.20$^{ 0.14}_{- 0.14}$ & 1.29$^{ 0.07}_{- 0.07}$ & 1.22$^{ 0.04}_{- 0.04}$ & 1.24$^{ 0.05}_{- 0.05}$ &6158$\pm$ 80& 3.5$^{ 0.5}_{- 0.5}$ &10.0$^{16.0}_{-16.0}$\\
   HAT-P-18& 5.5080&0.056&0.084$^{0.048}_{- 0.05}$ & 0.20$^{ 0.01}_{- 0.01}$ & 0.99$^{ 0.05}_{- 0.05}$ & 0.77$^{ 0.03}_{- 0.03}$ & 0.75$^{ 0.04}_{- 0.04}$ &4803$\pm$ 80& 0.5$^{ 0.5}_{- 0.5}$ &-\\
   HAT-P-19& 4.0088&0.047&0.067$^{0.042}_{- 0.04}$ & 0.29$^{ 0.02}_{- 0.02}$ & 1.13$^{ 0.07}_{- 0.07}$ & 0.84$^{ 0.04}_{- 0.04}$ & 0.82$^{ 0.05}_{- 0.05}$ &4990$\pm$130& 0.7$^{ 0.5}_{- 0.5}$ &-\\
   HAT-P-20& 2.8753&0.036&0.015$^{0.005}_{- 0.01}$ & 7.29$^{ 0.25}_{- 0.25}$ & 0.87$^{ 0.03}_{- 0.03}$ & 0.76$^{ 0.03}_{- 0.03}$ & 0.69$^{ 0.02}_{- 0.02}$ &4595$\pm$ 80& 2.1$^{ 0.5}_{- 0.5}$ &-\\
   HAT-P-22& 3.2122&0.041&0.016$^{0.009}_{- 0.01}$ & 2.15$^{ 0.08}_{- 0.08}$ & 1.08$^{ 0.06}_{- 0.06}$ & 0.92$^{ 0.04}_{- 0.04}$ & 1.04$^{ 0.04}_{- 0.04}$ &5302$\pm$ 80& 0.5$^{ 0.5}_{- 0.5}$ &-\\
   HAT-P-24& 3.3552&0.047&0.067$^{0.024}_{- 0.02}$ & 0.69$^{ 0.04}_{- 0.04}$ & 1.24$^{ 0.07}_{- 0.07}$ & 1.19$^{ 0.04}_{- 0.04}$ & 1.32$^{ 0.07}_{- 0.07}$ &6373$\pm$ 80&10.0$^{ 0.5}_{- 0.5}$ &20.0$^{16.0}_{-16.0}$\\
   HAT-P-25& 3.6528&0.047&0.032$^{0.022}_{- 0.02}$ & 0.57$^{ 0.03}_{- 0.03}$ & 1.19$^{ 0.08}_{- 0.06}$ & 1.01$^{ 0.03}_{- 0.03}$ & 0.96$^{ 0.05}_{- 0.04}$ &5500$\pm$ 80& 0.5$^{ 0.5}_{- 0.5}$ &-\\
   HAT-P-27& 3.0396&0.040&0 & 0.62$^{ 0.03}_{- 0.03}$ & 1.02$^{ 0.07}_{- 0.06}$ & 0.92$^{ 0.06}_{- 0.06}$ & 0.87$^{ 0.04}_{- 0.04}$ &5190$\pm$165& 0.6$^{ 0.7}_{- 0.4}$ &24.2$^{76.0}_{-44.5}$\\
   HAT-P-28& 3.2572&0.043&0.051$^{0.033}_{- 0.03}$ & 0.63$^{ 0.04}_{- 0.04}$ & 1.21$^{ 0.10}_{- 0.10}$ & 1.02$^{ 0.05}_{- 0.05}$ & 1.10$^{ 0.09}_{- 0.07}$ &5680$\pm$ 90& 0.2$^{ 0.5}_{- 0.5}$ &-\\
   HAT-P-29& 5.7232&0.067&0.095$^{0.047}_{- 0.05}$ & 0.78$^{ 0.06}_{- 0.06}$ & 1.11$^{ 0.11}_{- 0.11}$ & 1.21$^{ 0.05}_{- 0.05}$ & 1.22$^{ 0.13}_{- 0.07}$ &6087$\pm$ 88& 3.9$^{ 0.5}_{- 0.5}$ &-\\
    HAT-P-3& 2.8997&0.039&0 & 0.60$^{ 0.02}_{- 0.02}$ & 0.90$^{ 0.04}_{- 0.05}$ & 0.93$^{ 0.04}_{- 0.05}$ & 0.83$^{ 0.03}_{- 0.04}$ &5185$\pm$ 80& 0.5$^{ 0.5}_{- 0.5}$ &-\\
   HAT-P-35& 3.6467&0.050&0.025$^{0.018}_{- 0.02}$ & 1.05$^{ 0.04}_{- 0.04}$ & 1.33$^{ 0.10}_{- 0.10}$ & 1.24$^{ 0.05}_{- 0.05}$ & 1.44$^{ 0.08}_{- 0.08}$ &6096$\pm$ 88& 0.5$^{ 0.5}_{- 0.5}$ &-\\
   HAT-P-36& 1.3273&0.024&0.063$^{0.032}_{- 0.03}$ & 1.84$^{ 0.10}_{- 0.10}$ & 1.26$^{ 0.07}_{- 0.07}$ & 1.02$^{ 0.05}_{- 0.05}$ & 1.10$^{ 0.06}_{- 0.06}$ &5560$\pm$100& 3.6$^{ 0.5}_{- 0.5}$ &-\\
   HAT-P-37& 2.7974&0.038&0.058$^{0.038}_{- 0.04}$ & 1.17$^{ 0.11}_{- 0.11}$ & 1.18$^{ 0.08}_{- 0.08}$ & 0.93$^{ 0.04}_{- 0.04}$ & 0.88$^{ 0.06}_{- 0.04}$ &5500$\pm$100& 3.1$^{ 0.5}_{- 0.5}$ &-\\
   HAT-P-39& 3.5439&0.051&0 & 0.60$^{ 0.10}_{- 0.10}$ & 1.57$^{ 0.11}_{- 0.81}$ & 1.40$^{ 0.05}_{- 0.05}$ & 1.62$^{ 0.08}_{- 0.06}$ &6430$\pm$100&12.7$^{ 0.5}_{- 0.5}$ &-\\
   HAT-P-40& 4.4572&0.061&0 & 0.62$^{ 0.08}_{- 0.08}$ & 1.73$^{ 0.06}_{- 0.06}$ & 1.51$^{ 0.04}_{- 0.51}$ & 2.21$^{ 0.06}_{- 0.06}$ &6080$\pm$100& 6.9$^{ 0.5}_{- 0.5}$ &-\\
   HAT-P-41& 2.6940&0.043&0 & 0.80$^{ 0.10}_{- 0.10}$ & 1.69$^{ 0.08}_{- 0.05}$ & 1.42$^{ 0.05}_{- 0.05}$ & 1.68$^{ 0.06}_{- 0.04}$ &6390$\pm$100&19.6$^{ 0.5}_{- 0.5}$ &-\\
   HAT-P-49& 2.6915&0.044&0 & 1.73$^{ 0.21}_{- 0.21}$ & 1.41$^{ 0.13}_{- 0.08}$ & 1.54$^{ 0.05}_{- 0.05}$ & 1.83$^{ 0.14}_{- 0.08}$ &6820$\pm$ 52&16.0$^{ 0.5}_{- 0.5}$ &-\\
    HAT-P-8& 3.0763&0.045&0 & 1.29$^{ 0.05}_{- 0.05}$ & 1.50$^{ 0.07}_{- 0.07}$ & 1.28$^{ 0.04}_{- 0.04}$ & 1.58$^{ 0.08}_{- 0.06}$ &6200$\pm$ 80&11.5$^{ 0.5}_{- 0.5}$ &-9.7$^{ 9.0}_{- 7.7}$\\
    HAT-P-9& 3.9229&0.053&0 & 0.78$^{ 0.09}_{- 0.09}$ & 1.40$^{ 0.06}_{- 0.06}$ & 1.28$^{ 0.13}_{- 0.13}$ & 1.32$^{ 0.07}_{- 0.07}$ &6350$\pm$150&11.9$^{ 1.0}_{- 1.0}$ &16.0$^{ 8.0}_{- 8.0}$\\
     HATS-2& 1.3541&0.023&0 & 1.35$^{ 0.15}_{- 0.15}$ & 1.17$^{ 0.03}_{- 0.03}$ & 0.88$^{ 0.04}_{- 0.04}$ & 0.90$^{ 0.02}_{- 0.02}$ &5227$\pm$ 95& 1.5$^{ 0.5}_{- 0.5}$ &-\\
  HD 149026& 2.8759&0.043&0 & 0.36$^{ 0.02}_{- 0.02}$ & 0.65$^{ 0.06}_{- 0.05}$ & 1.29$^{ 0.06}_{- 0.05}$ & 1.37$^{ 0.12}_{- 0.08}$ &6160$\pm$ 50& 6.0$^{ 0.5}_{- 0.5}$ &12.0$^{ 7.0}_{- 7.0}$\\
  HD 209458& 3.5247&0.047&0 & 0.69$^{ 0.02}_{- 0.02}$ & 1.36$^{ 0.01}_{- 0.01}$ & 1.13$^{ 0.03}_{- 0.02}$ & 1.16$^{ 0.01}_{- 0.02}$ &6065$\pm$ 50& 4.5$^{ 0.5}_{- 0.5}$ &-4.4$^{ 1.4}_{- 1.4}$\\
   HD 97658& 9.4909&0.080&0.064$^{0.061}_{- 0.04}$ & 0.02$^{ 0.00}_{- 0.00}$ & 0.21$^{ 0.02}_{- 0.01}$ & 0.75$^{ 0.03}_{- 0.03}$ & 0.70$^{ 0.04}_{- 0.03}$ &5119$\pm$ 80& 0.5$^{ 0.5}_{- 0.5}$ &-\\
  Kepler-12& 4.4380&0.056&0.000$^{0.010}_{- 0.00}$ & 0.43$^{ 0.04}_{- 0.04}$ & 1.70$^{ 0.03}_{- 0.03}$ & 1.17$^{ 0.05}_{- 0.05}$ & 1.48$^{ 0.03}_{- 0.03}$ &5947$\pm$100& 0.8$^{ 0.5}_{- 0.5}$ &-\\
  Kepler-14& 6.7901&0.081&0.035$^{0.018}_{- 0.02}$ & 8.41$^{ 0.29}_{- 0.29}$ & 1.14$^{ 0.07}_{- 0.05}$ & 1.51$^{ 0.04}_{- 0.04}$ & 2.05$^{ 0.11}_{- 0.08}$ &6395$\pm$ 60& 7.9$^{ 1.0}_{- 1.0}$ &-\\
  Kepler-17& 1.4857&0.027&0.000$^{0.011}_{- 0.00}$ & 2.48$^{ 0.10}_{- 0.10}$ & 1.33$^{ 0.04}_{- 0.04}$ & 1.16$^{ 0.06}_{- 0.06}$ & 1.05$^{ 0.03}_{- 0.03}$ &5781$\pm$ 85& 6.0$^{ 2.0}_{- 2.0}$ &-\\
   Kepler-4& 3.2135&0.046&0 & 0.08$^{ 0.01}_{- 0.01}$ & 0.36$^{ 0.02}_{- 0.02}$ & 1.22$^{ 0.05}_{- 0.09}$ & 1.49$^{ 0.07}_{- 0.08}$ &5857$\pm$120& 2.1$^{ 1.0}_{- 1.0}$ &-\\
  Kepler-40& 6.8735&0.081&0 & 2.18$^{ 0.34}_{- 0.34}$ & 1.17$^{ 0.04}_{- 0.04}$ & 1.48$^{ 0.06}_{- 0.06}$ & 2.13$^{ 0.06}_{- 0.06}$ &6510$\pm$100& 9.0$^{ 2.0}_{- 2.0}$ &-\\
  Kepler-41& 1.8556&0.029&0 & 0.49$^{ 0.07}_{- 0.07}$ & 0.84$^{ 0.03}_{- 0.03}$ & 0.94$^{ 0.09}_{- 0.09}$ & 0.97$^{ 0.03}_{- 0.03}$ &5660$\pm$100& 4.5$^{ 1.5}_{- 1.5}$ &-\\
 Kepler-412& 1.7209&0.030&0.004$^{0.009}_{- 0.00}$ & 0.94$^{ 0.87}_{- 0.87}$ & 1.32$^{ 0.04}_{- 0.04}$ & 1.17$^{ 0.09}_{- 0.09}$ & 1.29$^{ 0.04}_{- 0.04}$ &5750$\pm$ 90& 5.0$^{ 1.0}_{- 1.0}$ &-\\
  Kepler-43& 3.0241&0.045&0.000$^{0.025}_{- 0.00}$ & 3.23$^{ 0.18}_{- 0.18}$ & 1.20$^{ 0.06}_{- 0.06}$ & 1.32$^{ 0.09}_{- 0.09}$ & 1.42$^{ 0.07}_{- 0.07}$ &6041$\pm$143& 5.5$^{ 1.5}_{- 1.5}$ &-\\
  Kepler-44& 3.2467&0.045&0.000$^{0.021}_{- 0.00}$ & 1.02$^{ 0.07}_{- 0.07}$ & 1.24$^{ 0.07}_{- 0.07}$ & 1.19$^{ 0.10}_{- 0.10}$ & 1.52$^{ 0.09}_{- 0.09}$ &5757$\pm$134& 4.0$^{ 2.0}_{- 2.0}$ &-\\
   Kepler-5& 3.5485&0.051&0 & 2.12$^{ 0.08}_{- 0.08}$ & 1.43$^{ 0.05}_{- 0.05}$ & 1.37$^{ 0.04}_{- 0.06}$ & 1.79$^{ 0.04}_{- 0.06}$ &6297$\pm$ 60& 4.8$^{ 1.0}_{- 1.0}$ &-\\
   Kepler-6& 3.2347&0.046&0 & 0.67$^{ 0.03}_{- 0.03}$ & 1.32$^{ 0.03}_{- 0.03}$ & 1.21$^{ 0.04}_{- 0.04}$ & 1.39$^{ 0.02}_{- 0.03}$ &5647$\pm$ 44& 3.0$^{ 1.0}_{- 1.0}$ &-\\
   Kepler-7& 4.8855&0.062&0 & 0.44$^{ 0.04}_{- 0.04}$ & 1.48$^{ 0.05}_{- 0.05}$ & 1.35$^{ 0.00}_{- 0.13}$ & 1.84$^{ 0.05}_{- 0.07}$ &5933$\pm$ 44& 4.2$^{ 0.5}_{- 0.5}$ &-\\
  Kepler-77& 3.5788&0.045&0 & 0.43$^{ 0.03}_{- 0.03}$ & 0.96$^{ 0.02}_{- 0.02}$ & 0.95$^{ 0.04}_{- 0.04}$ & 0.99$^{ 0.02}_{- 0.02}$ &5520$\pm$ 60& 1.5$^{ 1.0}_{- 1.0}$ &-\\
   Kepler-8& 3.5225&0.048&0 & 0.59$^{ 0.10}_{- 0.10}$ & 1.42$^{ 0.06}_{- 0.06}$ & 1.21$^{ 0.07}_{- 0.06}$ & 1.49$^{ 0.05}_{- 0.06}$ &6213$\pm$150&10.5$^{ 0.7}_{- 0.7}$ & 5.0$^{ 7.0}_{- 7.0}$\\
OGLE2-TR-L9& 2.4855&0.041&0 & 4.57$^{ 1.51}_{- 1.51}$ & 1.61$^{ 0.04}_{- 0.04}$ & 1.52$^{ 0.08}_{- 0.08}$ & 1.53$^{ 0.04}_{- 0.04}$ &6933$\pm$ 58&39.3$^{ 0.4}_{- 0.4}$ &-\\
    Qatar-1& 1.4200&0.023&0 & 1.09$^{ 0.09}_{- 0.09}$ & 1.16$^{ 0.04}_{- 0.04}$ & 0.85$^{ 0.03}_{- 0.03}$ & 0.82$^{ 0.03}_{- 0.03}$ &4861$\pm$125& 2.1$^{ 0.8}_{- 0.8}$ &-8.4$^{ 7.1}_{- 7.1}$\\
     TrES-2& 2.4706&0.036&0 & 1.20$^{ 0.05}_{- 0.05}$ & 1.22$^{ 0.04}_{- 0.04}$ & 0.98$^{ 0.06}_{- 0.06}$ & 1.00$^{ 0.03}_{- 0.03}$ &5850$\pm$ 50& 2.0$^{ 1.5}_{- 1.5}$ &-9.0$^{12.0}_{-12.0}$\\
     TrES-4& 3.5539&0.051&0 & 0.93$^{ 0.08}_{- 0.08}$ & 1.78$^{ 0.09}_{- 0.09}$ & 1.39$^{ 0.06}_{- 0.06}$ & 1.82$^{ 0.07}_{- 0.06}$ &6200$\pm$ 75& 9.5$^{ 1.0}_{- 1.0}$ & 7.3$^{ 4.6}_{- 4.6}$\\
     TrES-5& 1.4822&0.025&0 & 1.78$^{ 0.08}_{- 0.08}$ & 1.21$^{ 0.02}_{- 0.02}$ & 0.89$^{ 0.02}_{- 0.02}$ & 0.87$^{ 0.01}_{- 0.01}$ &5171$\pm$ 36& 3.8$^{ 0.4}_{- 0.4}$ &-\\
    WASP-10& 3.0928&0.038&0.051$^{0.008}_{- 0.01}$ & 3.19$^{ 0.12}_{- 0.12}$ & 1.08$^{ 0.02}_{- 0.02}$ & 0.79$^{ 0.02}_{- 0.03}$ & 0.70$^{ 0.01}_{- 0.01}$ &4675$\pm$100& 3.0$^{ 3.0}_{- 3.0}$ &-\\
   WASP-103& 0.9255&0.020&0 & 1.49$^{ 0.10}_{- 0.10}$ & 1.53$^{ 0.07}_{- 0.05}$ & 1.22$^{ 0.04}_{- 0.04}$ & 1.44$^{ 0.05}_{- 0.03}$ &6110$\pm$160&10.6$^{ 0.9}_{- 0.9}$ &-\\
    WASP-11& 3.7225&0.044&0 & 0.54$^{ 0.05}_{- 0.05}$ & 0.91$^{ 0.06}_{- 0.03}$ & 0.80$^{ 0.03}_{- 0.02}$ & 0.74$^{ 0.04}_{- 0.03}$ &4800$\pm$100& 0.5$^{ 0.2}_{- 0.2}$ &-\\
    WASP-13& 4.3530&0.054&0 & 0.48$^{ 0.06}_{- 0.06}$ & 1.39$^{ 0.04}_{- 0.06}$ & 1.09$^{ 0.05}_{- 0.05}$ & 1.51$^{ 0.03}_{- 0.04}$ &5826$\pm$100& 5.0$^{ 5.0}_{- 5.0}$ &-\\
    WASP-16& 3.1186&0.042&0 & 0.84$^{ 0.03}_{- 0.03}$ & 1.01$^{ 0.08}_{- 0.06}$ & 1.00$^{ 0.03}_{- 0.03}$ & 0.95$^{ 0.06}_{- 0.05}$ &5700$\pm$150& 3.0$^{ 1.0}_{- 1.0}$ &-4.2$^{11.0}_{-13.9}$\\
    WASP-18& 0.9415&0.020&0.008$^{0.001}_{- 0.00}$ &10.20$^{ 0.34}_{- 0.34}$ & 1.27$^{ 0.06}_{- 0.04}$ & 1.22$^{ 0.03}_{- 0.03}$ & 1.22$^{ 0.07}_{- 0.05}$ &6400$\pm$100&11.0$^{ 1.5}_{- 1.5}$ & 4.0$^{ 5.0}_{- 5.0}$\\
    WASP-19& 0.7888&0.016&0.005$^{0.004}_{- 0.00}$ & 1.13$^{ 0.04}_{- 0.04}$ & 1.39$^{ 0.03}_{- 0.03}$ & 0.93$^{ 0.02}_{- 0.02}$ & 0.99$^{ 0.02}_{- 0.02}$ &5500$\pm$100& 4.0$^{ 2.0}_{- 2.0}$ & 4.6$^{ 5.2}_{- 5.2}$\\
    WASP-21& 4.3225&0.052&0 & 0.30$^{ 0.01}_{- 0.01}$ & 1.07$^{ 0.06}_{- 0.06}$ & 1.01$^{ 0.03}_{- 0.03}$ & 1.06$^{ 0.04}_{- 0.04}$ &5800$\pm$100& 1.5$^{ 0.6}_{- 0.6}$ &-\\
    WASP-22& 3.5327&0.047&0.023$^{0.012}_{- 0.01}$ & 0.56$^{ 0.10}_{- 0.10}$ & 1.12$^{ 0.04}_{- 0.04}$ & 1.10$^{ 0.30}_{- 0.30}$ & 1.13$^{ 0.03}_{- 0.03}$ &6000$\pm$100& 3.5$^{ 0.6}_{- 0.6}$ &22.0$^{16.0}_{-16.0}$\\
    WASP-23& 2.9444&0.037&0.000$^{0.062}_{- 0.00}$ & 0.87$^{ 0.09}_{- 0.09}$ & 0.96$^{ 0.05}_{- 0.05}$ & 0.78$^{ 0.13}_{- 0.12}$ & 0.77$^{ 0.03}_{- 0.05}$ &5150$\pm$100& 2.2$^{ 0.3}_{- 0.3}$ &-\\
    WASP-24& 2.3412&0.037&0 & 1.09$^{ 0.04}_{- 0.04}$ & 1.30$^{ 0.04}_{- 0.04}$ & 1.18$^{ 0.03}_{- 0.03}$ & 1.33$^{ 0.03}_{- 0.03}$ &6075$\pm$100& 7.0$^{ 0.9}_{- 0.9}$ &-4.7$^{ 4.0}_{- 4.0}$\\
    WASP-25& 3.7648&0.047&0 & 0.58$^{ 0.04}_{- 0.04}$ & 1.22$^{ 0.06}_{- 0.05}$ & 1.00$^{ 0.03}_{- 0.03}$ & 0.92$^{ 0.04}_{- 0.04}$ &5750$\pm$100& 3.0$^{ 1.0}_{- 1.0}$ &14.6$^{ 6.7}_{- 6.7}$\\
    WASP-29& 3.9227&0.046&0.030$^{0.050}_{- 0.03}$ & 0.24$^{ 0.02}_{- 0.02}$ & 0.79$^{ 0.06}_{- 0.04}$ & 0.82$^{ 0.03}_{- 0.03}$ & 0.81$^{ 0.04}_{- 0.04}$ &4800$\pm$150& 1.5$^{ 0.6}_{- 0.6}$ &-\\
    WASP-32& 2.7187&0.039&0 & 3.45$^{ 0.14}_{- 0.14}$ & 1.10$^{ 0.04}_{- 0.04}$ & 1.07$^{ 0.05}_{- 0.05}$ & 1.09$^{ 0.03}_{- 0.03}$ &6140$\pm$ 95& 3.9$^{ 0.4}_{- 0.5}$ &10.5$^{ 6.4}_{- 5.9}$\\
    WASP-34& 4.3177&0.052&0.038$^{0.012}_{- 0.01}$ & 0.58$^{ 0.03}_{- 0.03}$ & 1.22$^{ 0.11}_{- 0.08}$ & 1.01$^{ 0.07}_{- 0.07}$ & 0.93$^{ 0.12}_{- 0.12}$ &5700$\pm$100& 1.4$^{ 0.6}_{- 0.6}$ &-\\
    WASP-35& 3.1616&0.043&0 & 0.72$^{ 0.06}_{- 0.06}$ & 1.32$^{ 0.03}_{- 0.03}$ & 1.07$^{ 0.02}_{- 0.02}$ & 1.09$^{ 0.02}_{- 0.02}$ &5990$\pm$ 80& 2.4$^{ 0.6}_{- 0.6}$ &-\\
    WASP-36& 1.5374&0.026&0 & 2.27$^{ 0.09}_{- 0.09}$ & 1.27$^{ 0.03}_{- 0.03}$ & 1.02$^{ 0.03}_{- 0.03}$ & 0.94$^{ 0.02}_{- 0.02}$ &5881$\pm$136& 3.2$^{ 1.3}_{- 1.3}$ &-\\
    WASP-37& 3.5775&0.045&0 & 1.79$^{ 0.17}_{- 0.17}$ & 1.16$^{ 0.07}_{- 0.06}$ & 0.93$^{ 0.12}_{- 0.12}$ & 1.00$^{ 0.05}_{- 0.05}$ &5800$\pm$150& 2.4$^{ 1.6}_{- 1.6}$ &-\\
    WASP-38& 6.8719&0.076&0.028$^{0.003}_{- 0.00}$ & 2.71$^{ 0.10}_{- 0.10}$ & 1.09$^{ 0.02}_{- 0.02}$ & 1.23$^{ 0.04}_{- 0.04}$ & 1.35$^{ 0.02}_{- 0.02}$ &6180$\pm$ 50& 7.5$^{ 0.0}_{- 0.3}$ & 7.5$^{ 4.7}_{- 6.1}$\\
    WASP-39& 4.0553&0.049&0 & 0.28$^{ 0.03}_{- 0.03}$ & 1.27$^{ 0.04}_{- 0.04}$ & 0.93$^{ 0.03}_{- 0.03}$ & 0.90$^{ 0.02}_{- 0.02}$ &5400$\pm$150& 1.4$^{ 0.6}_{- 0.6}$ &-\\
     WASP-4& 1.3382&0.023&0 & 1.22$^{ 0.05}_{- 0.05}$ & 1.34$^{ 0.02}_{- 0.03}$ & 0.91$^{ 0.05}_{- 0.05}$ & 0.91$^{ 0.02}_{- 0.02}$ &5500$\pm$150& 2.2$^{ 0.8}_{- 0.8}$ & 4.0$^{34.0}_{-43.0}$\\
    WASP-41& 3.0524&0.040&0 & 0.93$^{ 0.06}_{- 0.06}$ & 1.20$^{ 0.06}_{- 0.06}$ & 0.95$^{ 0.09}_{- 0.09}$ & 0.90$^{ 0.05}_{- 0.05}$ &5450$\pm$150& 1.6$^{ 1.1}_{- 1.1}$ &-\\
    WASP-42& 4.9817&0.055&0 & 0.50$^{ 0.03}_{- 0.03}$ & 1.06$^{ 0.05}_{- 0.05}$ & 0.88$^{ 0.09}_{- 0.08}$ & 0.85$^{ 0.04}_{- 0.04}$ &5200$\pm$150& 2.7$^{ 0.4}_{- 0.4}$ &-\\
    WASP-43& 0.8135&0.014&0 & 1.78$^{ 0.10}_{- 0.10}$ & 0.93$^{ 0.07}_{- 0.09}$ & 0.58$^{ 0.05}_{- 0.05}$ & 0.60$^{ 0.03}_{- 0.04}$ &4400$\pm$200& 4.0$^{ 0.4}_{- 0.4}$ &-\\
    WASP-44& 2.4238&0.035&0 & 0.89$^{ 0.06}_{- 0.06}$ & 1.14$^{ 0.11}_{- 0.11}$ & 0.95$^{ 0.03}_{- 0.03}$ & 0.93$^{ 0.07}_{- 0.06}$ &5410$\pm$510& 3.2$^{ 0.9}_{- 0.9}$ &-\\
    WASP-45& 3.1261&0.041&0 & 1.01$^{ 0.05}_{- 0.05}$ & 1.16$^{ 0.28}_{- 0.14}$ & 0.91$^{ 0.06}_{- 0.06}$ & 0.94$^{ 0.09}_{- 0.07}$ &5140$\pm$200& 2.3$^{ 0.7}_{- 0.7}$ &-\\
    WASP-46& 1.4304&0.024&0 & 2.10$^{ 0.09}_{- 0.09}$ & 1.31$^{ 0.05}_{- 0.05}$ & 0.96$^{ 0.03}_{- 0.03}$ & 0.92$^{ 0.03}_{- 0.03}$ &5620$\pm$160& 1.9$^{ 1.2}_{- 1.2}$ &-\\
    WASP-47& 4.1591&0.052&0 & 1.14$^{ 0.06}_{- 0.06}$ & 1.15$^{ 0.04}_{- 0.02}$ & 1.08$^{ 0.04}_{- 0.04}$ & 1.15$^{ 0.03}_{- 0.02}$ &5400$\pm$100& 3.0$^{ 0.6}_{- 0.6}$ &-\\
    WASP-48& 2.1436&0.034&0 & 0.98$^{ 0.09}_{- 0.09}$ & 1.67$^{ 0.08}_{- 0.08}$ & 1.19$^{ 0.04}_{- 0.04}$ & 1.75$^{ 0.07}_{- 0.07}$ &5920$\pm$150& 2.4$^{ 0.6}_{- 0.6}$ &-\\
    WASP-49& 2.7817&0.038&0 & 0.38$^{ 0.03}_{- 0.03}$ & 1.11$^{ 0.05}_{- 0.05}$ & 0.94$^{ 0.08}_{- 0.08}$ & 0.98$^{ 0.03}_{- 0.03}$ &5600$\pm$150& 0.9$^{ 0.3}_{- 0.3}$ &-\\
     WASP-5& 1.6284&0.027&0 & 1.62$^{ 0.06}_{- 0.06}$ & 1.14$^{ 0.10}_{- 0.04}$ & 1.01$^{ 0.04}_{- 0.04}$ & 1.03$^{ 0.07}_{- 0.04}$ &5880$\pm$150& 3.4$^{ 0.7}_{- 0.7}$ &12.1$^{ 8.0}_{-10.0}$\\
    WASP-50& 1.9551&0.029&0.009$^{0.011}_{- 0.01}$ & 1.47$^{ 0.09}_{- 0.09}$ & 1.15$^{ 0.05}_{- 0.05}$ & 0.89$^{ 0.08}_{- 0.07}$ & 0.84$^{ 0.03}_{- 0.03}$ &5400$\pm$100& 2.6$^{ 0.5}_{- 0.5}$ &-\\
    WASP-52& 1.7498&0.027&0 & 0.46$^{ 0.02}_{- 0.02}$ & 1.27$^{ 0.03}_{- 0.03}$ & 0.87$^{ 0.03}_{- 0.03}$ & 0.79$^{ 0.02}_{- 0.02}$ &5000$\pm$100& 2.5$^{ 1.0}_{- 1.0}$ &24.0$^{17.0}_{- 9.0}$\\
    WASP-54& 3.6936&0.050&0.067$^{0.033}_{- 0.03}$ & 0.63$^{ 0.03}_{- 0.03}$ & 1.65$^{ 0.09}_{- 0.08}$ & 1.21$^{ 0.03}_{- 0.03}$ & 1.83$^{ 0.09}_{- 0.08}$ &6100$\pm$100& 4.0$^{ 0.8}_{- 0.8}$ &-\\
    WASP-55& 4.4656&0.053&0 & 0.57$^{ 0.04}_{- 0.04}$ & 1.30$^{ 0.05}_{- 0.03}$ & 1.01$^{ 0.04}_{- 0.04}$ & 1.06$^{ 0.03}_{- 0.02}$ &5900$\pm$100& 3.1$^{ 1.0}_{- 1.0}$ &-\\
    WASP-56& 4.6171&0.056&0 & 0.61$^{ 0.04}_{- 0.04}$ & 1.09$^{ 0.04}_{- 0.03}$ & 1.11$^{ 0.02}_{- 0.02}$ & 1.11$^{ 0.03}_{- 0.02}$ &5600$\pm$100& 1.5$^{ 0.9}_{- 0.9}$ &-\\
    WASP-57& 2.8390&0.039&0 & 0.68$^{ 0.05}_{- 0.05}$ & 0.92$^{ 0.02}_{- 0.01}$ & 0.95$^{ 0.03}_{- 0.03}$ & 0.84$^{ 0.07}_{- 0.16}$ &5600$\pm$100& 3.7$^{ 1.3}_{- 1.3}$ &-\\
    WASP-58& 5.0172&0.056&0 & 0.89$^{ 0.07}_{- 0.07}$ & 1.37$^{ 0.20}_{- 0.20}$ & 0.94$^{ 0.10}_{- 0.10}$ & 1.17$^{ 0.13}_{- 0.13}$ &5800$\pm$150& 2.8$^{ 0.9}_{- 0.9}$ &-\\
    WASP-59& 7.9196&0.070&0.100$^{0.042}_{- 0.04}$ & 0.86$^{ 0.04}_{- 0.04}$ & 0.78$^{ 0.07}_{- 0.07}$ & 0.72$^{ 0.04}_{- 0.04}$ & 0.61$^{ 0.04}_{- 0.04}$ &4650$\pm$150& 2.3$^{ 1.5}_{- 1.5}$ &-\\
     WASP-6& 3.3610&0.043&0.054$^{0.018}_{- 0.02}$ & 0.52$^{ 0.02}_{- 0.02}$ & 1.22$^{ 0.05}_{- 0.05}$ & 0.93$^{ 0.02}_{- 0.02}$ & 0.87$^{ 0.03}_{- 0.04}$ &5450$\pm$100& 1.4$^{ 1.0}_{- 1.0}$ &-11.0$^{18.0}_{-14.0}$\\
    WASP-61& 3.8559&0.051&0 & 2.05$^{ 0.08}_{- 0.08}$ & 1.24$^{ 0.03}_{- 0.03}$ & 1.22$^{ 0.07}_{- 0.07}$ & 1.36$^{ 0.03}_{- 0.03}$ &6250$\pm$150&10.3$^{ 0.5}_{- 0.5}$ &-\\
    WASP-62& 4.4120&0.057&0 & 0.56$^{ 0.04}_{- 0.04}$ & 1.39$^{ 0.06}_{- 0.06}$ & 1.25$^{ 0.05}_{- 0.05}$ & 1.28$^{ 0.05}_{- 0.05}$ &6230$\pm$ 80& 8.7$^{ 0.4}_{- 0.4}$ &-\\
    WASP-63& 4.3781&0.057&0 & 0.38$^{ 0.03}_{- 0.03}$ & 1.43$^{ 0.10}_{- 0.06}$ & 1.32$^{ 0.05}_{- 0.05}$ & 1.88$^{ 0.10}_{- 0.06}$ &5550$\pm$100& 2.8$^{ 0.5}_{- 0.5}$ &-\\
    WASP-64& 1.5733&0.027&0 & 1.27$^{ 0.08}_{- 0.08}$ & 1.27$^{ 0.04}_{- 0.04}$ & 1.00$^{ 0.03}_{- 0.03}$ & 1.06$^{ 0.03}_{- 0.03}$ &5550$\pm$150& 3.4$^{ 0.8}_{- 0.8}$ &-\\
    WASP-66& 4.0861&0.055&0 & 2.31$^{ 0.13}_{- 0.13}$ & 1.39$^{ 0.09}_{- 0.09}$ & 1.30$^{ 0.07}_{- 0.07}$ & 1.75$^{ 0.09}_{- 0.09}$ &6600$\pm$150&13.4$^{ 0.9}_{- 0.9}$ &-\\
    WASP-67& 4.6144&0.052&0 & 0.42$^{ 0.03}_{- 0.03}$ & 1.40$^{ 0.30}_{- 0.20}$ & 0.87$^{ 0.04}_{- 0.04}$ & 0.87$^{ 0.04}_{- 0.04}$ &5200$\pm$100& 2.1$^{ 0.4}_{- 0.4}$ &-\\
    WASP-72& 2.2167&0.037&0 & 1.41$^{ 0.06}_{- 0.06}$ & 1.01$^{ 0.12}_{- 0.18}$ & 1.33$^{ 0.04}_{- 0.04}$ & 1.71$^{ 0.16}_{- 0.09}$ &6250$\pm$100& 6.0$^{ 0.7}_{- 0.7}$ &-\\
    WASP-78& 2.1752&0.036&0 & 0.88$^{ 0.08}_{- 0.08}$ & 1.70$^{ 0.11}_{- 0.11}$ & 1.33$^{ 0.09}_{- 0.09}$ & 2.20$^{ 0.12}_{- 0.12}$ &6100$\pm$150& 7.2$^{ 0.8}_{- 0.8}$ &-\\
    WASP-80& 3.0679&0.034&0.000$^{0.070}_{- 0.00}$ & 0.55$^{ 0.04}_{- 0.04}$ & 0.95$^{ 0.03}_{- 0.03}$ & 0.58$^{ 0.05}_{- 0.05}$ & 0.57$^{ 0.02}_{- 0.02}$ &4145$\pm$100& 3.5$^{ 0.3}_{- 0.3}$ &-\\
       XO-1& 3.9415&0.049&0 & 0.92$^{ 0.08}_{- 0.08}$ & 1.21$^{ 0.05}_{- 0.04}$ & 1.03$^{ 0.06}_{- 0.06}$ & 0.93$^{ 0.04}_{- 0.03}$ &5750$\pm$ 75& 1.1$^{ 0.1}_{- 0.1}$ &-\\
       XO-5& 4.1878&0.051&0 & 1.15$^{ 0.09}_{- 0.09}$ & 1.03$^{ 0.06}_{- 0.05}$ & 1.00$^{ 0.03}_{- 0.03}$ & 1.05$^{ 0.05}_{- 0.04}$ &5510$\pm$ 44& 1.8$^{ 0.5}_{- 0.5}$ &-\\

\end{longtable}
\end{document}